\documentclass[preprint]{aastex}

\usepackage{url}
\usepackage{amsmath}
\usepackage{longtable}
\usepackage{rotating}
\usepackage{afterpage}
\usepackage{hyperref}

\usepackage{natbib}
\usepackage[usenames]{color}

\newcommand{\Msun}[0]{\mbox{M}_\odot}

\newcommand{\Ha}[0]{\mbox{H}\alpha}

\newcommand{\eqn}[1]{Equation~(\ref{#1})}	
\newcommand{\fig}[1]{Figure~\ref{#1}}		
\newcommand{\sect}[1]{\S\ref{#1}}			
\newcommand{\tabl}[1]{Table~\ref{#1}}		

\newcommand{\diskfit}[0]{{\sc diskfit}}
\newcommand{\hsim}[0]{{\sc hsim}}
\newcommand{\HSIM}[0]{{\sc hsim}}
\newcommand{\rtoh}[0]{{\sc ramses2hsim}}

\newcommand{\SIMA}[0]{{\tt NUTFB}}

\newcommand{\INFID}[0]{{\tt IN-FID}}

\newcommand{\INHALF}[0]{{\tt IN-$\Delta$1}}
\newcommand{\INDOUB}[0]{{\tt IN-$\Delta$6}}

\newcommand{\OUTFID}[0]{{\tt OUT-FID}}

\newcommand{\OUTTINY}[0]{{\tt OUT-T2}}
\newcommand{\OUTSHORT}[0]{{\tt OUT-T4}}
\newcommand{\OUTLONG}[0]{{\tt OUT-T60}}
\newcommand{\OUTHALF}[0]{{\tt OUT-$\Delta$1}}
\newcommand{\OUTDOUB}[0]{{\tt OUT-$\Delta$6}}

\definecolor{Red}{rgb}{0.65,0.08,0.05}
\definecolor{Blue}{rgb}{0,0.08,0.65}
\definecolor{Green}{rgb}{0,0.65,0.08}

\begin{document}

\title{Simulating gas kinematic studies of high-redshift galaxies with the HARMONI Integral Field Spectrograph}
\author{Richardson, Mark L. A.\altaffilmark{0,1,2,3}, Routledge, L.\altaffilmark{1}, Thatte,
  N.\altaffilmark{1}, Tecza, M.\altaffilmark{1}, Houghton, Ryan C. W.\altaffilmark{1}, Pereira-Santaella,
  M.\altaffilmark{1,4}, Rigopoulou, D.\altaffilmark{1}}
\altaffiltext{0}{Mark.Richardson@queensu.ca}
\altaffiltext{1}{Department of Physics, University of Oxford, Denys Wilkinson Building, Keble Road, Oxford, UK, OX1 3RH}
\altaffiltext{2}{Arthur B. McDonald Canadian Astroparticle Physics Research Institute, 64 Bader Lane, Kingston, ON, Canada, K7L 3N6}
\altaffiltext{3}{Department of Physics, Engineering Physics, and Astronomy, Queen's University, 64 Bader Lane, Kingston, ON, Canada, K7L 3N6}
\altaffiltext{4}{Centro de Astrobiolog\'ia (CSIC-INTA), Ctra. de Ajalvir, Km 4, 28850, Torrej\'on de Ardoz, Madrid, Spain}

\setcounter{footnote}{0}

\begin{abstract}

We present simulated observations of gas kinematics in galaxies formed
in 10 pc resolution cosmological simulations with the hydrodynamical + N-body code
{\sc ramses}, using the new \rtoh\ pipeline with the simulated observing
pipeline (\HSIM) for the ELT HARMONI IFU spectrograph.
We post-process the galaxy's gas kinematics and H$\alpha$ line emission for each simulation cell, and integrate the emission to produce an extinction-corrected input cube. 
We then simulate observations of the input cube with
HARMONI, for a range of exposure times, spatial sampling, and spectral resolution. We analyze the mock observations to recover galaxy properties 
such as its kinematics and compare with the known simulation values. We investigate the cause of biases between the `real' and `observed' kinematic values, demonstrating the
sensitivity of the inferred rotation curve to knowledge of the
instrument's point spread function. 


\end{abstract}
\maketitle
\setcounter{footnote}{0}

\section{Introduction}
The cosmic star formation history is recognized to peak
around $1 < z < 3$ \citep{Madau14}.  However, the physical mechanism that drives the star
formation in galaxies is yet unclear. It has been proposed that continuous
accretion of cold gas from the intergalactic medium could drive
intense episodes of star formation in massive highly unstable high
redshift galaxies \citep{Dekel09,Dave12}. Alternatively, starburst
events could be fuelled through gas-rich major mergers
\citep[e.g.,][]{Hayward14,Riechers14}. Higher resolution observations will be crucial to make further progress on this question.

The physical processes associated with star formation can be better understood via resolved studies
of galaxy gas kinematics using integral field spectroscopy observations of nebular line emission, such as the H$\alpha$ emission expected from star-forming
regions (e.g., SINS -- \citealt{ForsterSchreiber09,ForsterSchreiber18}; KMOS3D -- \citealt{Wisnioski15,Wisnioski19}; KROSS --
\citealt{Stott2016}; MOSDEF -- \citealt{Kriek15,Price16,Price20}; ZFIRE -- \citealt{Straatman17}). The Extremely Large Telescope (ELT) will be able to continue these studies efficiently at moderate redshifts. Thus, it is essential that we understand the capabilities of this new telescope and its first-light instruments.

HARMONI is the first-light, adaptive optics (AO) assisted,
near-infrared integral field spectrograph (IFS) for the ELT \citep{Thatte16}. HARMONI will greatly increase the observable spatial resolution for galaxies near the peak epoch of galactic star formation. This will allow for comparison with local galaxies, and give insight into how galaxies have changed over cosmic time. As it is a new instrument and telescope, early understanding of its observational features are paramount. To facilitate the community being introduced to the HARMONI instrument, a dedicated instrument simulation pipeline, \HSIM\  \citep{Zieleniewski15}, was created to expedite the proposal, observation, and analysis pipeline, and demonstrate to the community what HARMONI and ELT capabilities will be.

While new observational facilities are progressing, high resolution simulations are probing star forming processes at a range of redshifts to better constrain their gas physics and yield better predictions for future observations. To be more effective at making inferences based on the results of galaxy simulations, the simulation community has made efforts to make mock observations. Mock observations also allow for a direct quantification of observational biases that occur in recovering physical characteristics of real galaxies. \citet{Kaviraj17} made mock luminosity functions using simulated galaxies to determine whether their galaxy populations resembled those in reality. \citet{Hirschmann17} used their detailed simulations of galaxies from high redshift to present day to connect changes in emission line diagnostic ratios to the metallicity, ionization parameter, star formation, and feedback from active galactic nuclei. \citet{Keating20} post-processed their simulation of a Milky Way-like galaxy to determine their resulting CO to H$_2$ conversion factor. Finally, \citet{Guidi18} have made a mock CALIFA \citep{Sanchez12} catalog of simulated galaxies' resolved nebular (and stellar) emission, and they have released that data publicly for community comparisons between their simulated galaxy catalog and real galaxies in CALIFA. Indeed there is significant value in releasing tools to the community to facilitate further comparisons between simulations and observation.

There are three methods for {\em creating} source galaxies at high redshift to study the impact of telescope and instrument on the data analysis and interpretation.  One is to artificially redshift nearby objects for which high spatial and spectral resolution, high signal-to-noise data are available  (c.f. \citealt{Garcialorenzo2019}).  However, this method ignores the morphological, kinematic and dynamical differences between galaxies at $1 \lesssim z \lesssim 3$ and those at $z = 0$, which are known to be substantial.  The second method is to create ad-hoc galaxies whose properties (e.g. clump size and distribution) match those of observed high redshift objects (c.f. \citealt{Zieleniewski15b}, \citealt{Kendrew16}); the data for the spatially resolved properties come from observations of strong gravitationally-lensed systems.  However, these ad-hoc objects are not designed to be dynamically stable, and thus may not correctly represent real high-z objects.  The third, preferred, method is to use cosmological simulations that forward propagate primordial density fluctuations consistent with observations of the cosmic microwave background, creating individual galaxies at high spatial resolution, whose kinematic, morphology and dynamical properties are consistent with observed ensemble properties of the population at the corresponding redshift (c.f., \citealt{Kendrew16,Kaviraj17,Guidi18}). As the input physics (e.g. star formation laws) for the simulation is well understood, the resulting objects provide robust, reliable, mock galaxies consistent with physical laws and cosmological evolution models.  Here we propose a method of post-processing these mock galaxies, computing gas emission line intensities to get realistic model galaxy observations with self-consistent kinematics and dynamics. Future papers will explore the potential of using comparisons between real observations and predicted observations to discriminate between models, thus constraining the physical parameters (e.g. slope of the stellar initial mass function) used in the cosmological simulations.

One purpose of this paper is to present the \rtoh\ software pipeline \citep{Richardson20} that can be used to convert simulated galaxies from the adaptive mesh refinement (AMR) hydrodynamical code RAMSES \citep{Teyssier02} into input emission cubes for \HSIM. Both the simulations
and observations reach comparable spatial resolutions of order
10--100\,pc. While \rtoh\ is made available as an \href{https://github.com/mlarichardson/ramses2hsim}{online resource}\footnote{Repository available at \href{https://github.com/mlarichardson/ramses2hsim}{https://github.com/mlarichardson/ramses2hsim}}, here we demonstrate its utility in exploring the complete pipeline from simulated galaxy to observed IFS cube and its analysis. In particular, the key purpose of this paper is to showcase through this pipeline the impacts of certain HARMONI instrument characteristics and AO performance on the recovered physical characteristics of the galaxy. In the process, we will highlight new features of the HARMONI instrument, in particular information about the point-spread function (PSF) of the telescope and detector. A commonly discussed bias in recovering galaxy characteristics is beam smearing \citep[e.g.,][]{Davies11,Genzel14,diTeodoro15,Burkert16,ForsterSchreiber18,Simons19}, where the velocity gradient is smoothed out due to limited spatial resolution, which is expected to appear here. These past works attempt to correct for this, typically with good success.

The structure of this paper is as follows. Section \sect{method}
describes how the input and output cubes are created, and analysed to
infer the galaxy kinematics. As examples of converting simulated galaxies into inferred emission flux are limited in the literature, we describe the process in significant detail so that other simulation codes can implement an equivalent tool. 
Such a pipeline can be instrumental in connecting future simulated galaxy observables with both observations and fundamental galaxy processes. Section \sect{results} presents the
results for integral field
spectroscopy observations, for a range of exposure times, and various estimates of
the instrument's point spread function (PSF).  Finally, section
\sect{conclusions} discusses the results and presents the conclusions.

\section{Numerical Methods}\label{method}
Here we give a detailed discussion of the simulation used for this work and the \rtoh\ pipeline made available that converts the simulation dataset into an input cube for the HARMONI simulation. Note that the simulation is used as a method
to make a cosmologically self-consistent galaxy, albeit not
necessarily representative of a typical object at that cosmic epoch. Quantifying the impact of the observing setup and characteristics, e.g., the AO PSF, on the observations and their interpretation compared to the underlying simulation quantities is the focus of this paper.

\subsection{Simulation setup}\label{sims}
For the present study we used the largest galaxy from the {\tt NUTFB} simulation in the {\tt NUT}  simulation suite
\citep{Powell11}. {\tt NUT} is a collection of zoom cosmological
simulations using the adaptive mesh refinement (AMR) code {\tt RAMSES} \citep{Teyssier02}, where each simulation used identical initial conditions, centred on the
formation of a $3.6\times 10^{11}$ M$_\odot$ DM halo at $z=0$. The simulation suite varied one or more of: final redshift $z_{\rm f}$,
star formation model, stellar feedback model, resolution, radiative
transfer, and magnetic fields. High-resolution simulations ended at $z_{\rm
  f}=3$ due to the computation expense to evolve them lower. Depending on the model, the $z = 3$ halo has a mass of $1.2 -
1.3\times 10^{11}$ M$_\odot$, with a galaxy stellar mass of $5.3\times
10^9 - 1.3 \times 10^{10}$ M$_{\odot}$, and a galaxy gas fraction of
$12 - 31$\%. \tabl{tab_Rs} lists several details of the galaxy used in this study.

The galaxy was taken from the lowest redshift \SIMA\ output at $z_{\tt f}=3$. For the purpose of this work, we are demonstrating a method for approximating the gas-phase emission from the galaxy, and determining how such a galaxy would be observed with the HARMONI instrument, including the impact of the AO PSF on the observations. While our focus is not on the galaxy being representative, we include details of its stellar formation and feedback prescriptions for transparency.

{\small
\begin{longtable}{lccccccccccc}
\caption[Simulated Galaxy Properties]{Simulated Galaxy Properties}\label{tab_Rs} \vspace{-8pt} 

\\ \hline
\multicolumn{1}{l}{\textbf{Sim}$^{\rm a}$} &
\multicolumn{1}{c}{$\Delta x^{\rm b}$} &
\multicolumn{1}{c}{r$_{200}^{\rm c}$} &
\multicolumn{1}{c}{M$_{\rm DM,200}^{\rm d}$} &
\multicolumn{1}{c}{r$_{\operatorname{c-h}}^{\rm e}$} &
\multicolumn{1}{c}{r$_{\tt e}^{\rm f}$} &                
\multicolumn{1}{c}{M$_*^{\rm g}$} &
\multicolumn{1}{c}{M$_{\rm gas}^{\rm h}$}  &
\multicolumn{1}{c}{M$_{\rm total}^{\rm i}$} &
\multicolumn{1}{c}{dM$_*$/dt$^{\rm j}$} &
\multicolumn{1}{c}{dM$_*$/dt$^{\rm k}$} &
\multicolumn{1}{c}{$i^{\rm l}$} \\
 & & & & & & & & & (inst) & (avg) & \\\hline
\SIMA 	     & 11.9 & 43 & 1.3E11 &  1.7 & 0.37 & 1.28E10 & 1.8E9 & 2.06E10 & 3.27 & 2.53  & 51$^\circ$  \\ \hline 
\multicolumn{11}{l}{\ \ $^{\rm a}$Simulation name} \\
\multicolumn{11}{l}{\ \ $^{\rm b}$Simulation resolution (pc)} \\
\multicolumn{11}{l}{\ \ $^{\rm c}$Halo virial radius (kpc)} \\
\multicolumn{11}{l}{\ \ $^{\rm d}$Halo dark matter mass ($\Msun$)} \\
\multicolumn{11}{l}{\ \ $^{\rm e}$Galaxy cold-hot gas transition radius (kpc)} \\
\multicolumn{11}{l}{\ \ $^{\rm f}$Galaxy stellar half-mass radius (kpc)} \\
\multicolumn{11}{l}{\ \ $^{\rm g}$Galaxy stellar mass within r$_{\operatorname{c-h}}$ ($\Msun$)} \\
\multicolumn{11}{l}{\ \ $^{\rm h}$Galaxy gas mass within r$_{\operatorname{c-h}}$ ($\Msun$)} \\
\multicolumn{11}{l}{\ \ $^{\rm i}$Galaxy total mass (incl. DM) within r$_{\operatorname{c-h}}$ ($\Msun$)} \\
\multicolumn{11}{l}{\ \ $^{\rm j}$Galaxy instantaneous star formation rate  ($\Msun$ yr$^{-1}$)} \\
\multicolumn{11}{l}{\ \ $^{\rm k}$Galaxy star formation rate over 50 Myr ($\Msun$ yr$^{-1}$)} \\
\multicolumn{11}{l}{\ \ $^{\rm l}$Galaxy angular momentum vector inclination with respect to the line-of-sight.}
\end{longtable}
}

The star formation prescription in {\tt NUTFB} depends only on the gas density, with an average value given by
(see \citet{Rasera2006} and references therein),
\begin{equation}\label{SFR}
\frac{dM_*}{dt} = \epsilon \frac{M_{\rm gas}}{t_{\rm ff}} = 57 \left(\frac{\epsilon}{0.01}\right)\left(\frac{V}{\mbox{kpc}^3}\right)\left(\frac{n_{\rm H}}{400\mbox{\ cm}^{-3}}\right)^{3/2} \left(\frac{X}{0.76}\right)^{-1}\mbox{\ M}_\odot\mbox{\ yr}^{-1},
\end{equation}
where $M_* = dM_*/dt \times \Delta t$ is the average stellar mass created in a new star particle during the simulation time step $\Delta t$, $M_{\rm gas}$ is the gas mass in an AMR cell of volume $V$, and $X$ is the mass fraction of hydrogen, set to 0.76. Only gas with hydrogen number density, $n_{\rm H}$, larger than a threshold density, $n_0$, here set to $400$ cm$^{-3}$, forms stars over a freefall time, $t_{\rm ff}$ at a fixed efficiency, $\epsilon=0.008$. The star formation mass at any time follows a Poisson statistic with expectation value $M_*$ to be more stochastic.

The stellar feedback prescription in {\tt NUTFB} follows \citet{Dubois08}. This model includes energy injected from type-II supernovae (SNe-II). Once a star particle reaches an age of 10 Myr it injects $\eta_{\rm SN} \times 10^{50}$ erg M$_\odot^{-1}$ of specific internal energy, where $\eta_{\rm SN}$ is the mass fraction of a stellar population that undergoes a type-II supernova. {\tt NUTFB} uses a Salpeter initial mass function (IMF), thus $\eta_{\rm SN}$ = 0.106. This energy is split evenly between a thermal and kinetic component. The supernova also injects mass and metals corresponding to this mass fraction, and a yield of $y = 0.1$, respectively. The mass and energy are injected following the analytic Sedov-Taylor solution for a blast wave. The prescription includes a mass-loading of $\eta_{\rm W} = 10$, corresponding to the mass swept up inside the injection bubble of size $r_{\rm bubble} = 32$ pc.

\subsection{Constructing mock input cubes for \HSIM}\label{halpha}

We extracted a cubic region from \SIMA\ that is 14 kpc across and centred on the galaxy with a stellar truncation radius of 3.1 kpc, nearly double the transition radius from cool to hot gas, r$_{\operatorname{c-h}} = 1.7$ kpc. Each simulation cell in this volume was mapped to a uniform resolution `\textit{emission}' datacube
with spatial coordinate axes. We integrated the $\Ha$ emission and
metals from all simulation cells contributing to a given emission datacube
cell, and flux-averaged the simulation cell kinematics and
temperature. This was then further processed to construct an integral
field spectrograph `\textit{input}' datacube for {\tt HSIM}, with two
spatial dimensions and one wavelength dimension. Here we outline this process in detail and make the pipeline available online for others to use.

To construct the emission datacube, we took each native simulation gas
cell, described by a physical (x,y,z) position relative to the centre
of its host halo, then determined which cell in the emission datacube
it fell within. The emission datacube had a resolution of 24 pc, twice
coarser than the resolution of the native RAMSES output. We also looked at results for emission datacubes with half and double this fiducial resolution. For each simulation cell we determined the total $\Ha$ emission following \cite*{Kennicutt94}  (hereafter K94), where
ionizing photons, resulting from young, massive stars and therefore
directly related to the star formation rate, are reprocessed by the
neutral medium and some fraction are re-emitted as H$\alpha$ through a recombination cascade. Thus, for each cell we calculate what would be the instantaneous star formation rate given the simulation star formation model, and convert this to H$\alpha$ using the conversion: 
\begin{equation}\label{Ha_Kenn94}
L_{\rm H\alpha} = 1.26\times 10^{41} \left({\rm \frac{dM_*}{dt} }{\rm M}_\odot{\rm \,yr^{-1}}\right) \mbox{\ erg s}^{-1},
\end{equation}
\begin{equation}\label{Ha_SFR2}
L_{\rm H\alpha} = 7.2\times 10^{42} \left(\frac{\epsilon}{0.01}\right)\left(\frac{V}{\mbox{kpc}^3}\right)\left(\frac{n_{\rm H}}{400\mbox{\ cm}^{-3}}\right)^{3/2} \left(\frac{X}{0.76}\right)^{-1} \mbox{\ erg s}^{-1},
\end{equation}
where we have substituted in \eqn{SFR} for the star formation rate
from a single cell. 

Each cell in the emission datacube was assigned the total $\Ha$ flux and average metallicity of its overlapping simulation cells. Further, the flux-averaged velocity and local turbulence were also saved. The local turbulence, $\sigma_{\rm gas}$, follows those used by subsequent simulations in the {\tt NUT} suite, where the turbulence is estimated by the trace of the square of the velocity gradient tensor.

\begin{figure}[h!]
\centering
\includegraphics[width=0.9\columnwidth]{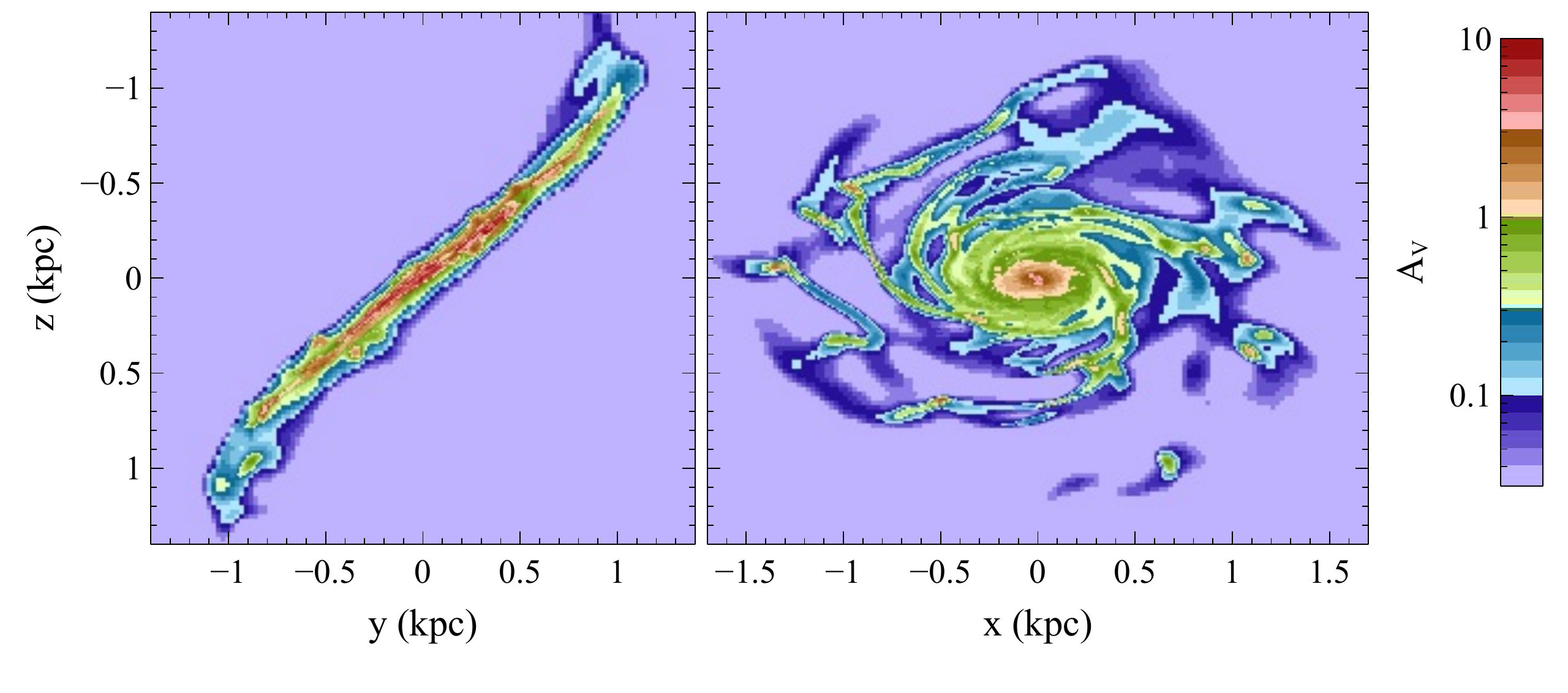}
\caption{\small Projected $A_{\rm V}$ maps along the $x$ (left) and $y$ (right) axes. The scaling is logarithmic in A$_{\rm V}$,
with each hue change representing 0.5 dex, and each brightness step representing 0.1 dex.}
\label{fig_ext}
\end{figure}

We then integrated through the emission datacube along a line of sight to produce the input datacube. We integrated along the simulation $y$-axis as the galaxy's inclination along this line of sight is close to 45$^\circ$. Integrating along the line of sight is necessary to properly account for dust extinction, where only foreground dust can extinct background emission. We estimated the extinction from dust for a given cell in the emission datacube by assuming it scales with the foreground column density of gas-phase metals, $\Sigma_{\rm Z}$, between that cell and the front side of the emission box. We consider an extinction coefficient $Q_\lambda
= Q_0 a/\lambda \simeq Q_0 a/\lambda_{\rm H\alpha}$
\citep{CarrollOstlie}, thus
\begin{equation}
A_{\rm V} = 1.086\left(\frac{3}{4\rho_{\rm d}}\right)f_{\rm d} \Sigma_{\rm Z} Q_0/\lambda_{\rm V}, 
\end{equation}
where dust grains of size $a$ follow a power law $dn/da \sim
a^{-3.5}$, over the range $a_1 = 0.005\ \mu m < a < a_2 =1\ \mu m$,  a
single dust grain has density $\rho_{\rm d} \simeq 3\,\mbox{g}\
\mbox{cm}^{-3}$, and a fraction $f_{\rm d}$ of gas-phase metals are
locked into dust. We assume a dust fraction of 8\%, slightly lower than seen in
observations (e.g. \citealt{Peeples14}), an $A_{\rm V} \lesssim
1$, similar to seen for local and $z\sim 1.5$ galaxies. \citep[e.g.,][]{Calzetti00,Erb06,Dominguez13,Kahre18} 
A given cell also has self-absorption by assuming the $\Ha$ emission and the dust mass are uniformly distributed.
Maps of the total galaxy extinction along different lines-of-sight
are shown in
\fig{fig_ext}, showing as expected increased extinction in the core of the galaxy and along its spiral arms. 

The input cube was originally constructed in a
frame at rest, with a velocity resolution of 10 km\,s$^{-1}$, or 0.22
\AA. Each emission cell then contributed to the input cube as a
Gaussian emission line, with total flux equal to the total $\Ha$
emission of that cell extincted by $e^{-A_{\rm V}/1.32}$, and width
$\sigma$, where $\sigma^2 = \sigma_{\rm gas}^2/3 + \sigma_{\rm th}^2$ and 
$\sigma_{\rm th} = \sqrt{2P/\rho}$ is the thermal width ($P$ is
the mean gas pressure in the cell, and $\rho$ is the gas density). Note that the local turbulence is typically much larger
than the thermal width. The two spatial axes are rescaled from
physical pc to angular milli-arc seconds using the angular diameter distance to the galaxy for the desired redshift.

The wavelength dimension, centred at $\Ha$ at rest, was then shifted
to account for the galaxy redshift. For this work we artificially place the
galaxy extracted from the $z=3$ snapshot at $z=1.44$, leaving the
galaxy unchanged in physical extent. Thus, we only change the wavelengths,
luminosity and angular diameter distance to the galaxy. We do this for
two reasons. The first is that we are making mock images of the $\Ha$
line, which is redshifted outside the wavelength coverage of HARMONI
at $z=3$. By assuming the galaxy is at $z=1.44$, the $\Ha$ line is
instead redshifted to the center of the H band. It was computationally
too expensive to evolve the entire cosmological simulation to
$z=1.44$, given the number of time steps required. Second, the
galaxy considered here has low star formation rates at its final redshift (see \tabl{tab_rotpeak}) and would be difficult to
detect at $z=3$. By placing the galaxy at lower $z$, we have better
SNR. Recall, the purpose of this work is not to showcase observations of highly realistic simulated galaxies. Instead, the purpose is to highlight the specifics of observing a galaxy with HARMONI on the ELT, including how physical characteristics of the galaxy is biased in the process. To convert to $z = 1.44$, we use the 7-year WMAP cosmological parameters
($H_0$, $\Omega_M$) = (70.4 km\,s$^{-1}$, 0.272) \citep{Komatsu2011}.
Since
the physical galaxy is not modified in this process, the redshift
$z=1.44$ is only used when redshifting the input cube to the H band.

\subsection{Observational properties of the input galaxy}\label{galprop}
Our fiducial input cube uses the {\tt NUTFB} simulation at
$z=3$, which has a 0.3 $L_*(z=3)$ galaxy \citep{Oesch10}. The
simulation has a resolution of 12 pc, but our fiducial cube has
a resolution of 24 pc for reasonable file size. We perform the
mock observations with the same physical galaxy but as if it was at
$z=1.44$, placing $\Ha$ in the centre of the H-band, and corresponding
to a compact 1.25 $L_*(z=1.44)$ galaxy \citep{Oesch10}. Due to
inefficient feedback in the {\tt NUTFB} simulation, the central bulge is
very compact,  causing our effective radius to be small compared to typical galaxies at this redshift \cite[e.g.,][]{Price16}, and a very fast rotation speed \cite[e.g.,][]{Ubler17}, peaking
at 480 km\,s$^{-1}$ at 60 pc. At $z=1.44$, the 24 pc resolution
corresponds to 2.78 mas resolution. The input cube's wavelength sampling is roughly 0.5 \AA, or 10 km\,s$^{-1}$ per pixel at
the observed H$\alpha$ wavelength of $1.6\,\mu {\rm m}$.

The measured line-of-sight rotation velocity is deprojected to infer the intrinsic rotation velocity of the galaxy. This deprojection
requires accurate knowledge of the galaxy's inclination to the
line-of-sight. While the difficulty in determining an inclination of an observed galaxy is well understood, we wish to highlight that even in a simulation where the characteristics of the galaxy are quantified exactly, bulk parameters of the galaxy, such as inclination, are very sensitive to the medium being measured and over what scale. For this galaxy, the inclination of gas on small scales (400 pc) is 53.8$^\circ$, while on larger scales (1.7 kpc) the inclination is 49.1$^\circ$. Stars on the 400 pc scale are more inclined at roughly 56.5$^\circ$, while young stars less than 5 Myrs old over the 1.7 kpc scale have roughly the same inclination as gas, with a value of 51$^\circ$. For this work we use the inclination of young stars on large scales, 51$^\circ$. The uncertainty in the derived rotation curve is dependent on the uncertainty in the  inclination. However, the variation in inclination
stemming from different media and scales is similar to
the uncertainty in a real observation, where the inclination has to be
derived from isophotal analysis of the photometry, assuming an
intrinsic round shape.

\subsection{Input and Output cube notations}\label{notations}
In this work we quantify how \HSIM\ recovers different galaxy characteristics for different observing setups, and how these characteristics relate to the intrinsic characteristics of the galaxy. Thus, we will relate the analysis of the output cubes to the native simulation values. However, the process of making the input cube for \HSIM\ may cause additional biases, which we attempt to quantify. Consequently, we vary both the pipeline for producing the input cube, and the \HSIM\ observing setup which produces a unique output cube. Here we discuss these different methods, and how we will refer to them. An overview is given in \tabl{tab_cubes}.

The fiducial input cube, notated as \INFID\ has a spaxel scale of 2.78 mas, and a wavelength sampling of 0.5
\AA. The input cubes \INHALF\
and \INDOUB\ are generated in the same manner as \INFID, except with
spaxel scales of 1.39 mas and 5.56 mas, respectively. For input and output cubes, it is vital to distinguish between
resolution and sampling.  Along the spectral direction, the input
cubes have a sampling of 0.5\,\AA, while the spectral resolution is
not well-defined, varying depending on the input line width, which is
typically $>$ 1\,\AA.  The fiducial output cube has a resolving power
R $\equiv \lambda /\Delta\lambda \approx 3500$ dictated by the HARMONI grating, and a wavelength sampling
of 2.78\,\AA, which is designed to be just Nyquist sampled for a
wavelength resolution of 5.56\,\AA . Along the two spatial axes, the
input cubes have a sampling of 2.78\,mas (except for the {\tt $\Delta
  1$} and {\tt $\Delta 6$} versions).  As this is already twice larger than the
  intrinsic cell size of the simulation, the resolution is also the
  same value, and resolution / sampling are used interchangeably when
  referring to the input cubes.  The output cube sampling is 10\,mas,
  which is sub-Nyquist, given the PSF FWHM of 15\,mas for these mock observations. Due to undersampling effects, the effective spatial resolution of the data is close to 20\,mas, and we use this value when comparing rotation curves derived from the input and output cubes.
  Note that the diffraction limited
  PSF width of the ELT is $\sim$10\,mas, but \HSIM\ includes contributions from the LTAO residual jitter,
  instrumental vibration \& wind-shake, and the spectrograph image quality degradation, achieving a PSF FWHM of 15\,mas. We discuss the impact of residual LTAO jitter in more detail in section \ref{psf}.

 \begin{longtable}{l cccc}
\caption[Input \& Output Cube Notations]{Input \& Output Cube Notations}\label{tab_cubes} \vspace{-8pt} \\
\hline
\multicolumn{5}{l}{\textbf{Input Cubes}} \\ \hline
\multicolumn{1}{l}{\textbf{Cube Name}} &
\multicolumn{1}{l}{\textbf{Spaxel Scale$^{\rm a}$}} &
\multicolumn{2}{l}{\textbf{Wavelength Sampling$^{\rm b}$}} \\ \hline
\INFID		 &  2.78 & 0.53 \\
\INHALF		 &  1.39 & 0.53 \\
\INDOUB		 &  5.56 & 0.53 \\
\hline
\multicolumn{5}{l}{\textbf{Output Cubes}} \\ \hline
\multicolumn{1}{l}{\textbf{Cube Name}} &
\multicolumn{1}{l}{\textbf{Input Cube}} &
\multicolumn{1}{l}{\textbf{Exposure Time}} &
\multicolumn{1}{l}{\textbf{Spaxel Scale$^{\rm a}$}} &
\multicolumn{1}{l}{\textbf{Spectral Resolution$^{\rm b}$}} \\ \hline
   \OUTFID		& \INFID 		& 20$\times$900s	& 10 & 5.56 \\
\OUTHALF	& \INHALF 	& 20$\times$900s	& 10 & 5.56 \\
\OUTDOUB	& \INDOUB 	& 20$\times$900s	& 10 & 5.56 \\
{\tt OUT-T2} & \INFID          & 2$\times$900s         & 10  & 5.56\\
\OUTSHORT	& \INFID 		& 4$\times$900s		& 10 & 5.56 \\
\OUTLONG	& \INFID 		& 60$\times$900s	& 10 & 5.56 \\
{\tt OUT-R7} & \INFID        & 20$\times$900s   & 10 & 2.78 \\
\hline
\multicolumn{5}{l}{$^{\rm a}$mas \ \ $^{\rm b}$\AA}
\vspace{-30pt}
\end{longtable}
  
\subsection{\HSIM\ output cubes}\label{output}
Almost all input cubes were `observed' with the same \HSIM\ pipeline
setup, specifically using 20$\times$900\,s exposures with a spaxel scale of 10
mas and a spectral resolution of $\approx$3500, or 5.56\,\AA. The observations were 
done using the Laser Tomographic Adaptive Optics mode of HARMONI. This observation
produced output reduced cubes, which reconstituted the data cube, and
included additional Poisson noise from a sky subtraction using a
random realisation of the sky, as is typical for near-IR
observations.
While the input cubes had standard units of intensity, the output
cubes had units of electrons in each spaxel. We call the fiducial
output cube, or \OUTFID, the result of `observing' the \INFID\ input
cube with this setup. We also observed the \INFID\ input cube using a
setup of 2$\times$, 4$\times$ and 60$\times$900s exposures, with all other settings the same
as in \OUTFID, notated as {\tt OUT-T2}, \OUTSHORT\ and \OUTLONG,
respectively. We also observed \INFID\ with the R $\approx 7000$ grating, yielding the {\tt OUT-R7} output cube, which has double the spectral resolution compared to the fiducial output cube. All other output cubes are generated with the same setup
as \OUTFID\ for their corresponding input cube, as indicated in
\tabl{tab_cubes}.

\subsection{Analysis} 

\subsubsection{H$\alpha$ Fits}
The data cubes are 3-dimensional, with the wavelength as the first
dimension, and the second and third dimensions being spatial.  The
data were first masked based on the integrated signal to noise ratio
(SNR) per spaxel over all spectral pixels of the emission line, with a threshold of 7, so
as to ensure reliable data. This threshold was chosen as there were
some skyline residuals still in the data that, at lower SNR, could mimic
an emission line. Where it was not possible to fit the spectrum due to
low SNR, an average of the surrounding pixels was taken over a 2$\times$2
spaxel box, to increase SNR at the expense of spatial
resolution. Alternative strategies to exploit low SNR data were also
investigated, and are described below.

\begin{figure}[ht]
    \centering
    \includegraphics[width=\textwidth]{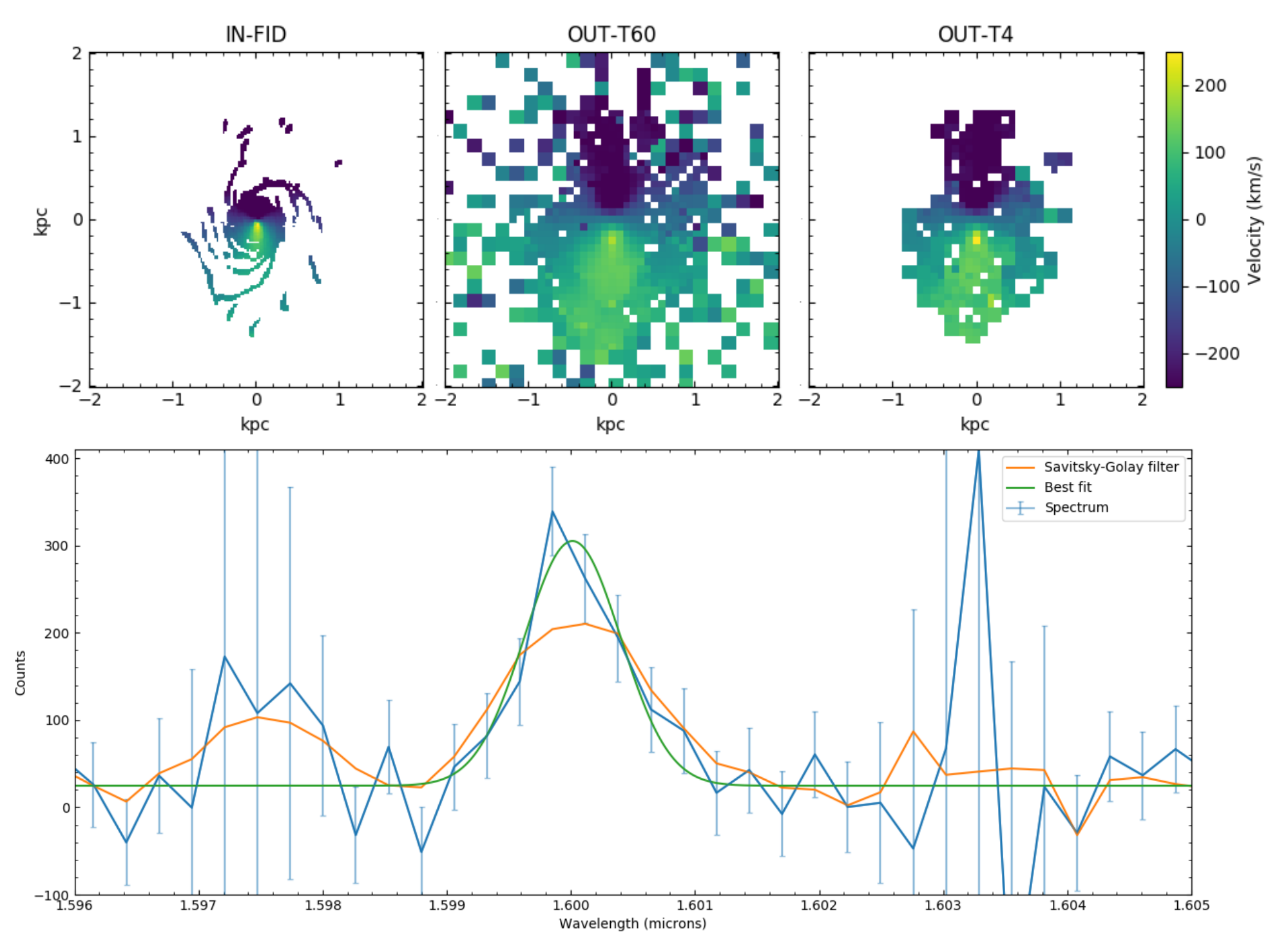}
    \caption{\small upper: 2D velocity maps for the \SIMA\ galaxy, showing the fitting algorithm applied to the input cube \INFID\ (left), high SNR output cube \OUTLONG\ (centre), and low SNR output cube {\tt OUT-T4} (right). White areas have no emission in the input cube. lower: An example of the fitting algorithm used, showing the spectrum (blue), the Savitsky-Golay filter (red) and the best fitting Gaussian (green). As can be seen, the correct peak has been identified, and the residual skyline noise seen around 1.603\,$\mu$m has been ignored. }
    \label{fig:vel-maps}
\end{figure}

The \hsim\ data cube is in wavelength space, however, given the
simulated galaxy's redshift of 1.44, wavelengths can be converted
to recession velocities, using the velocity zero point to be the H$\alpha$ nominal wavelength
of 1.60\,$\mu$m for this redshift. To determine the central wavelength
of the H$\alpha$ line for each spaxel a Gaussian fit was applied along the spectral
axis, incorporating the entire wavelength range contained within the
datacube, between 1.595\,$\mu$m and 1.606\,$\mu$m. This fit had 4 free
parameters, for central value, amplitude, width and a constant
background. To obtain initial guesses for the parameters, a
Savitzky--Golay filter was applied to the data with an 11\,\AA\ box to
smooth the noise, then the maximum of the smoothed data was found, and
taken to be the initial guess for the amplitude of the Gaussian, with
the corresponding wavelength taken as the guess for the central value
of the emission line. A non-linear least squares fitting method from
\verb|SciPy| was used, where the uncertainties incorporated into the
fit were taken to be the variance returned in the reduced cube from
\verb|HSIM|. The reduced cube is sky subtracted, assuming a perfect
knowledge of the sky flux, however still contains the associated
Poisson noise since a different instance of the background noise is
used in the sky subtraction. From this fit, we determine the line of
sight velocity of the galaxy at each spaxel or 2$\times$2 spaxel box, as well
as the dispersion from the width of the Gaussian function. \fig{fig:vel-maps} shows the 2D velocity maps returned from this fit
on 3 cubes for the \SIMA\ galaxy: an input cube, a high SNR cube, and a low SNR cube, along
with the fit applied to a single spaxel, including the data returned
from the simulation, the Savitzky-Golay filter and the returned best
fit spectrum. 

After this fitting algorithm was run, the following checks of the returned
parameters were implemented to ensure that the data used in further
analysis were reliable, and false positive fits, such as fits to noise
peaks, were not included. Data with extremely high dispersion (greater
than 600\,km\,s$^{-1}$) and those with a high residual background value
(greater than an absolute value of 50 counts) were rejected. Also
data with an uncertainty on the magnitude of the peak greater than the
absolute value of the peak were rejected, along with data where the
uncertainty on the dispersion was greater than 0.8 times the value of
the dispersion. The final checks were to remove data
where the maximum
amplitude is only twice that of the background, as such a line was
considered too weak to be reliable. These values were chosen to retain
the maximum amount of reliable data, whilst ensuring that very few
false positive fits were included.

We also considered alternative methods for determining the velocity
maps: evaluating the first moment or Voronoi binning. The first moment
along the spectral axis was computed, excluding the regions around the
sky lines. For Voronoi binning, bins were created with a SNR cutoff of
7, equivalent to the original fitting method. In both cases,
subsequent analysis yielded very similar results to the original ones.  The moment maps
fared a little worse when the SNR for a single spaxel was too low,
while there was no significant benefit to using Voronoi
binning. Therefore, we chose to use the analysis method described
above.

\subsubsection{DiskFit}
In order to incorporate the full 2D data, the software package
\diskfit\footnote{https://www.physics.queensu.ca/Astro/people/Kristine$\_$Spekkens/diskfit/} \citep{Spekkens07,Reese07,Sellwood10,Kuzio12,Sellwood15}
was used. This software takes an input data file containing
information about the 2D velocity profile of the galaxy, obtained from
(Gaussian) fitting to the emission line, along with parameters such as
initial guesses for galaxy centre, inclination and position angle, and the systemic
velocity of the galaxy. It fits circular rotation velocities to annuli
of specified width and spacing, and minimises $\chi^2$ for all annuli,
whilst varying the global parameters and rotation velocity for each
annulus.  As \diskfit\ does not have information about the emission
line flux morphology, it is unable to effectively deconvolve smearing
from the instrument PSF, so the PSF deconvolution feature was turned
off for the fitting.

We converted the already created 2D velocity maps
for the galaxy into the format required for
\diskfit, and then the fit was performed on these files. Due to
the size of some of the cubes, and in particular the input cubes, in
some cases only the central region of the data cube was used for computational efficiency, up to a radius of approximately 2\,kpc. Note that this truncation does not present any issues as the rotation curve could only be extracted for this central region with high SNR and not in the outer edges of the galaxy where the SNR significantly decreases.

The outputs from \diskfit\ depend on the parameters chosen to fit, which
in this case were centre of galaxy, systemic velocity and the
rotational velocity at each radius. The inclination to the line of
sight, and the position angle were held fixed, as they were known from
the simulation data. Uncertainties on the fit parameters were obtained
through a bootstrapping procedure inbuilt into \diskfit.  The
\diskfit\ output is rather sensitive to the centre coordinates,
so a two-step procedure was adopted. First, the outer radius was
constrained to 1\,kpc and the centre coordinates determined from a
fit. Then, with the centre coordinates held fixed at those values,
a second fit was carried out to a larger radius, limited by the SNR of
the data.  The 1.5\,kpc radius was chosen to match the region of
well-ordered rotation in the galaxy.

\section{Results and Discussion}\label{results}
The results of this work are highlighting the new features of the HARMONI instrument for various observing conditions and AO performance. These results are demonstrated through how well the HARMONI instrument recovers characteristics of the simulated galaxy as predicted by using {\tt HSIM}. Here we discuss first our
fiducial input and output cubes, \INFID\ and \OUTFID,
respectively. Then we discuss a rough convergence test using input cubes of varying spatial resolution (\INHALF\ and \INDOUB), and the resulting output cubes (\OUTHALF\ and \OUTDOUB). We then compare simulated observations with different observing times resulting in output cubes  \OUTTINY, \OUTSHORT\ and
\OUTLONG, as well as high spectral resolution with {\tt OUT-R7}. Finally, we present mock observations with elongated PSFs with 2:1 and 3:1 aspect ratios at various position angles. These elongated PSFs are typical of LTAO systems with off-axis natural guide stars, which is the case for HARMONI.

For a full description of the input and output cubes,
please see \sect{halpha}, \sect{output}, and \tabl{tab_cubes}. The
analysis of these cubes discusses three quantitative metrics, first
the peak SNR of the observation cube, second the radius at which the
rotation speed of the galaxy reaches a peak, and third the value of
the peak rotation.  We compare these with the values in the simulation,
and in the input cubes for {\tt HSIM}. This work highlights
biases that are introduced during observation and at what stage they occur, which can impact
measurements of e.g., the Tully-Fisher relation \citep{Tully77} and for
measuring the baryon fraction of disks (e.g., \citealt{Genzel17},
\citealt{Tiley16,Tiley19}). One such bias, beam smearing, is discussed and corrected for in many works \citep[e.g.,][]{Genzel06,Genzel08,Genzel11,Genzel14,Genzel17,Cresci09,Davies11,diTeodoro15,ForsterSchreiber18,Burkert16,Wuyts16,Tiley19,Simons19}. 

\subsection{Fiducial Input and Output Cubes}\label{fid}

\begin{figure}[h]
\centering
\includegraphics[scale=0.58]{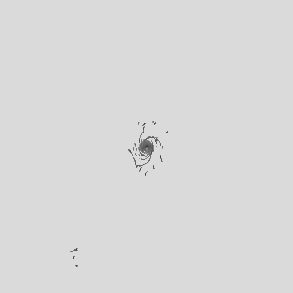}
\includegraphics[scale=0.521]{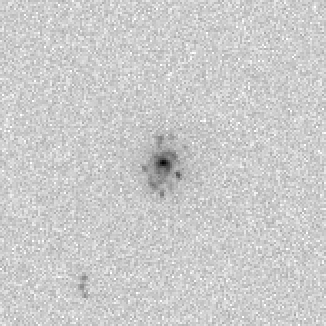}
\caption{\small Integrated H$\alpha$ intensity in the input cube \INFID\ (left) and resulting output cube \OUTFID\ (right). The region is 14\,kpc or  1.95$^{\prime\prime}$ across. The spatial sampling in the left image is 2.78\,mas, and in the right image is 10\,mas. Note that the brightness, contrast, and minimum and maximum values of the displayed images match between \INFID\ and \OUTFID.}
\label{fig_input}
\end{figure}

Projections of the \INFID\ and \OUTFID\ cubes are presented in
\fig{fig_input}. The output cube,
\OUTFID, was generated by observing \INFID\ using the {\tt HSIM}
pipeline with 20 900\,s exposures with the H+K R3500 grating, with a
spaxel scale of 10 mas. We find that the integrated $\Ha$ in the input cube
shows clear spiral arms where cold, star-forming gas is found, as well
as very dense but warm gas at the centre. We can also see a satellite
galaxy in the lower left which is in the process of accreting into the main galaxy, although it is far enough away to not be observed in our kinematic analysis of the main galaxy. In the
integrated output cube, the spatial resolution and sensitivity are
considerably worse, leading to difficulties in resolving the spiral
arms. For the output cube, the median integrated signal
to noise ratio (SNR) is 101 along the spectral axis for the central 60
mas diameter aperture, over 1.5954 $-$ 1.6060 $\mu$m.

The inferred star formation rates, as tabulated in
\tabl{tab_rotpeak}, are less than the intrinsic values from the simulation due to line-of-sight extinction within the
simulated galaxy.  For the output cubes, we
compute two numbers, the SFR deduced from the observed cube, summing
over a radius of r$_{\operatorname{c-h}}$, as tabulated in
\tabl{tab_Rs}, and an extinction corrected value, assuming A$_{\rm
  V}$=1 as seen for local and $z\sim 1$ galaxies \citep[eg.,][]{Calzetti00,Erb06,Dominguez13,Kahre18}. The AO PSF tends to have a
core-halo structure, and as the halo is very extended
($\gtrsim$100\,mas), only part of the source flux is contained within
the extraction aperture, reducing the inferred SFR compared to the
input cube value.

\begin{figure}[h]
\centering
\includegraphics[scale=0.50]{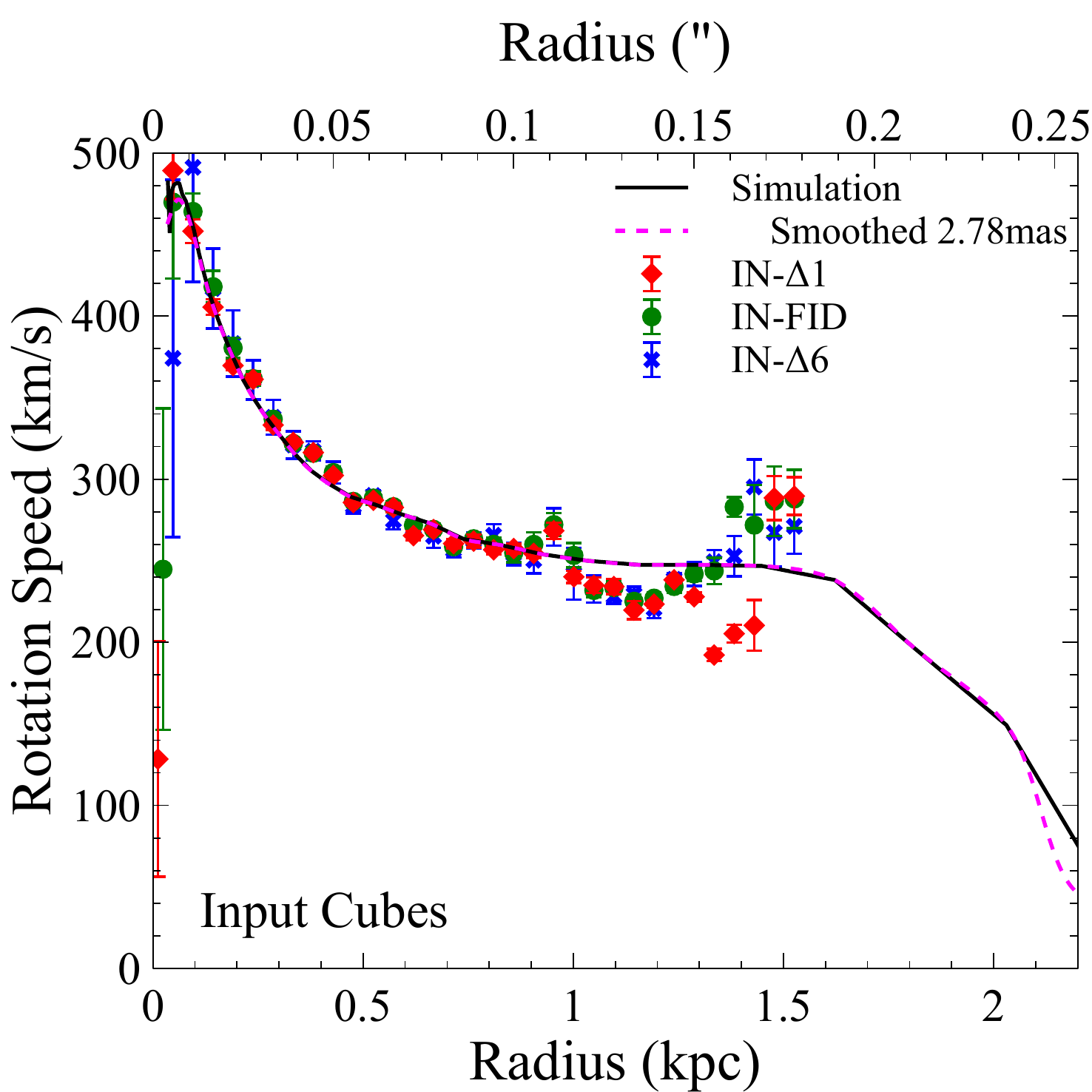}
\includegraphics[scale=0.50]{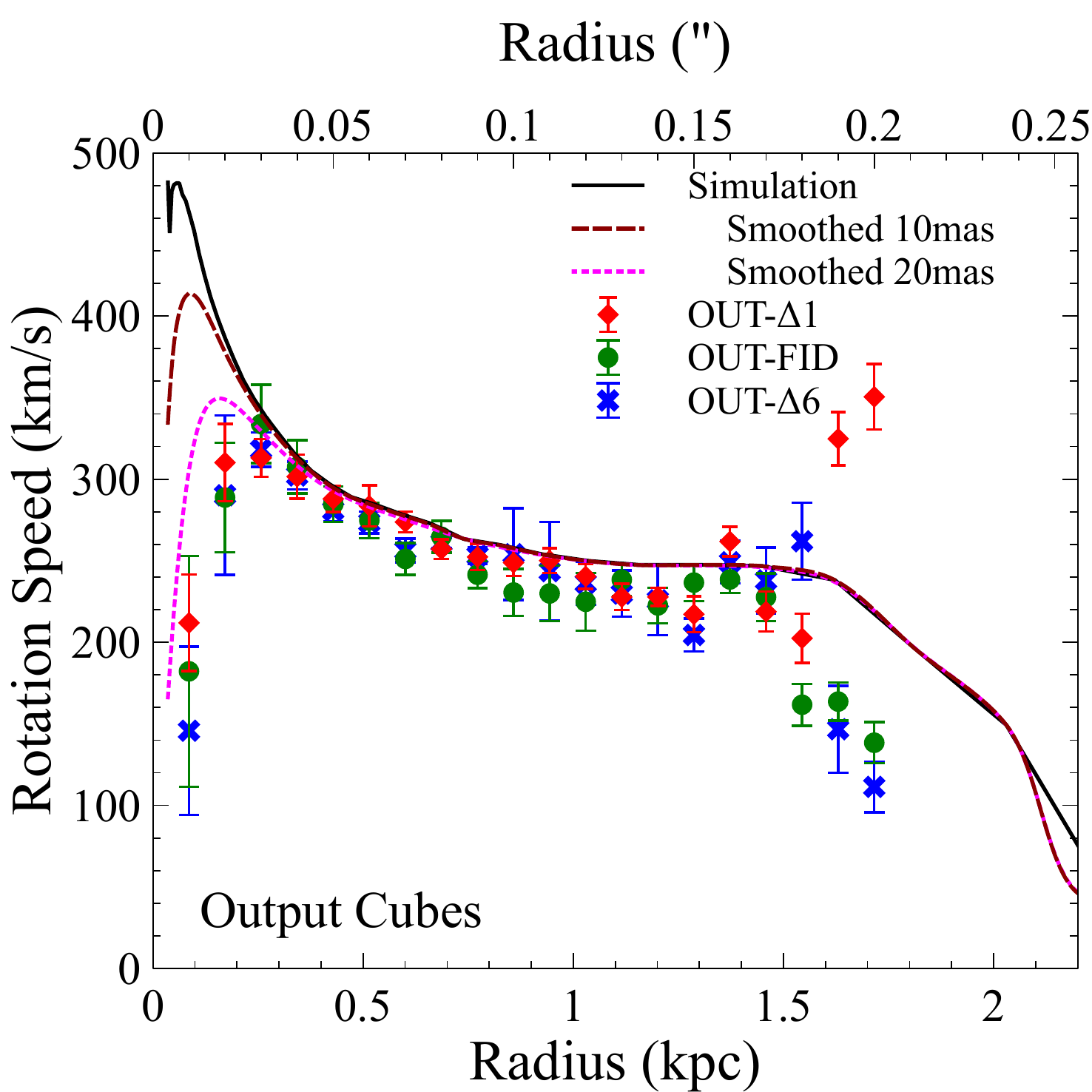}
\caption{\small Rotation curves of the galaxy for the input (left) and output (right) cubes for the {\tt FID} cube (green circles) with input spatial sampling of 2.78 mas, and $\Delta$1 (red diamonds) and $\Delta$6 (blue crosses), of spatial sampling 1.39 mas and 5.56 mas. We also include the unprocessed rotation curve from the simulation (black solid), as well as at the resolution of the input (magenta dashed), and the resolution of the output (dark red, dashed line) and twice this resolution (magenta dotted). Radius is taken in the plane of the disk.}
\label{fig_levels}
\end{figure}

In \fig{fig_levels} we show the results of fitting \diskfit\ 
to the {\tt IN-FID} and {\tt OUT-FID} cubes with profile plots of deprojected rotation velocity versus deprojected radius. We also include the raw rotation
curve of the simulation gas (which is consistent with circular
rotation) in the native resolution, and smoothed by convolving with a Gaussian of
FWHM 2.78 mas (input cube resolution), 10 mas (output cube spaxel
size), and 20 mas (twice the output cube spaxels).
Finally, we truncate the
rotation velocity measure where the emission signal has insufficient
signal to noise (spectrally integrated SNR $<$7), which is around 
1.8\,kpc for both input and output cubes.

As \diskfit\ uses a deprojected 2D fit, we account for projection effects skewing the shape of the galaxy, as well as pixels off the principle axis (determined from kinematics in the simulation) whose velocity includes components normal to the line of sight. The result is that in the input cube the
full simulation rotation curve is recovered very well, except at the
outskirts where the gas is not in ordered rotation, and its higher
dispersion increases the uncertainty in the \diskfit\ rotation speed
values. 

Analysis of the input cube yields a peak rotation of 476
$\pm$ 19 km\,s$^{-1}$ at 87 $\pm$ 19 pc, roughly consistent to 1$-\sigma$ with values
extracted from the simulation. However, the output cube suffers from lower spatial resolution, smearing out the central peak, suggesting a peak rotation speed of 334 $\pm$ 24 km\,s$^{-1}$ at 269
$\pm$ 48 pc. This is entirely in line with expectations, as the output
cube's resolution is limited by the instrument and AO PSF, with a
combined FWHM of $\approx$22\,mas for these mock observations. The observed rotation curve also suffers from a systematically underestimated value for the flat part of the rotation curve. We explore this in detail in \sect{psf}.

In \tabl{tab_rotpeak} we tabulate for both the input and output cubes
the peak rotation velocity and radius, and the integrated SFR inferred
from the $\Ha$ line emission, as well as the median integrated SNR
along the spectral axis within a 60 mas diameter aperture at the centre of the output cubes.

\begin{longtable}{l | ccc | ccc | c}
\caption[Rotation Characteristics Recovery]{Rotation Characteristics Recovery}\label{tab_rotpeak} \vspace{-8pt} \\
\hline
\multicolumn{1}{l}{\textbf{Simulation}} &
\multicolumn{2}{|c|}{Native $^{\rm a}$R$_{\rm p}$} &
\multicolumn{2}{c|}{Native $^{\rm b}$v$_{\rm p}$} &
\multicolumn{1}{c}{Native} &
\multicolumn{1}{c}{Native} \\
\multicolumn{1}{l}{ } &

\multicolumn{2}{|c|}{ } &
\multicolumn{2}{c|}{} &
\multicolumn{1}{c}{$^{\rm c}$SFR$_{\rm inst}$} &
\multicolumn{1}{c}{$^{\rm c}$SFR$_{50}$} \\ \hline
\multicolumn{1}{l}{Native res} 	&
    \multicolumn{2}{|c|}{60}	& 
    \multicolumn{2}{c|}{480} & 
    \multicolumn{1}{c}{3.27} & \multicolumn{1}{c}{2.53} \\ 
\multicolumn{1}{l}{\ \ at 10 mas}		& 
    \multicolumn{2}{|c|}{170}	& 
    \multicolumn{2}{c|}{400} & \multicolumn{1}{c}{-} & \multicolumn{1}{c}{-} \\
\multicolumn{1}{l}{\ \ at 20 mas}	&	
    \multicolumn{2}{|c|}{380}	& 
    \multicolumn{2}{c|}{330} & \multicolumn{1}{c}{-} & \multicolumn{1}{c}{-} \\ \hline

\multicolumn{1}{l}{\ } &
\multicolumn{3}{|c|}{Input Cubes} &
\multicolumn{3}{c|}{Output Cubes} &
\multicolumn{1}{c}{ \ } \\ 

\textbf{Analysis} &

$^{\rm a}$R$_{\rm p}$ &
$^{\rm b}$v$_{\rm p}$ &
$^{\rm c}$SFR &

$^{\rm a}$R$_{\rm p}$ &
$^{\rm b}$v$_{\rm p}$&
$^{\rm c}$SFR &

SNR \\ \hline
{\tt FID}	   	    & 87 $\pm$ 19		& 476 $\pm$ 19 	    & 1.49 	& 269 $\pm$ 48 	    & 334 $\pm$ 24 	& 1.41 (3.00) & 101 \\ 
$\Delta${\tt 1} 	& 71 $\pm$ 9		& 479 $\pm$ 20 	    & 0.96  & 232 $\pm$ 129	    & 314 $\pm$ 11 	& 0.93 (1.98) & 69 \\
$\Delta${\tt 6} 	& 100 $\pm$ 24		& 493 $\pm$ 10 	& 1.81  & 269 $\pm$ 56 	    & 319 $\pm$ 10 	& 1.78 (3.80) & 126 \\
{\tt OUT-T2}	& -	& -	&  - & 253 $\pm$ 286 	& 327 $\pm$ 106 & 1.36 (2.91) & 32 \\
{\tt OUT-T4}	& -	& -	&  - & 260 $\pm$ 48 	& 324 $\pm$ 13 	& 1.55 (3.31) & 45 \\
{\tt OUT-T60}	& -	& -	&  - & 254 $\pm$ 54 	& 318 $\pm$ 13 	& 1.46 (3.11) & 175 \\ 
{\tt OUT-R7}	& -	& -	&  - & 252 $\pm$ 59 	& 325 $\pm$ 11 	& 1.47 (3.15) & 148 \\
\hline
\multicolumn{8}{l}{$^{\rm a}$pc \ \ $^{\rm b}$km\,s$^{-1}$ \ \ $^{\rm
  c}$M$_{\odot}$ yr$^{-1}$; values is brackets assume A$_{\rm V}$=1.}
\end{longtable}

We constructed two additional input cubes using twice and half the
resolution of the fiducial input cube. We refer to these as cubes
\INHALF\ and \INDOUB\ respectively, as the resolutions are 1.89 and
5.56 mas, respectively. For the reduced output cube, the central
median spectrally integrated SNR is 69 for \OUTHALF\ and 126 for \OUTDOUB, each evaluated over a 60\,mas diameter aperture. The higher SNR
for \OUTDOUB\ is due to sampling fewer simulation cells per output
cube spaxel, which effectively smooths the data, lowering the
variance.  We plot the recovered rotation curves in \fig{fig_levels} for the input and output
cubes. The output cubes are generated using the same setup for \hsim\
as for the fiducial cube. The rotation curves for the input cube show convergence as they are
insensitive to the resolution of the cube at intermediate
distances.  At large scales where the spiral arms are small, the arms do not land on many spaxels at a fixed radius, and thus some spaxels are consistent with a lower rotation speed. The \INDOUB\ cube matches almost exactly the {\tt IN-FID} rotation curve, showing clear convergence at this level of spatial sampling. The analysis of the output cubes, which are all at 10 mas resolution, are equally insensitive to the resolution, and only the outskirts vary as larger scale turbulent gas is convolved into these scales.

We find that for the input cubes, we recover peak rotations of
479 $\pm$ 20 km\,s$^{-1}$ and 493 $\pm$ 10 km\,s$^{-1}$ at 71 $\pm$ 9
pc and 100 $\pm$ 24 pc for \INHALF\ and \INDOUB, respectively,
consistent with values extracted from the simulation. For the output cubes we
recover peak rotations of 314 $\pm$ 11 km\,s$^{-1}$ and 319 $\pm$ 10 km\,s$^{-1}$ at 232 $\pm$ 129 pc and 269 $\pm$ 56 pc for \OUTHALF\ and \OUTDOUB, respectively. The result as seen in past works \citep[e.g.,][]{Davies11,ForsterSchreiber18,Tiley20} is that for recovering a galaxy's peak velocity value and radius there is a trade-off between having more spatial resolution and having more SNR. Trying to correct for this beam smearing is difficult here where the velocity peak occurs at a radius roughly equal to the beam's half width at half max of $\approx$ 11 mas. 

\begin{figure}[t!]
  \centering
  \includegraphics[scale=0.34]{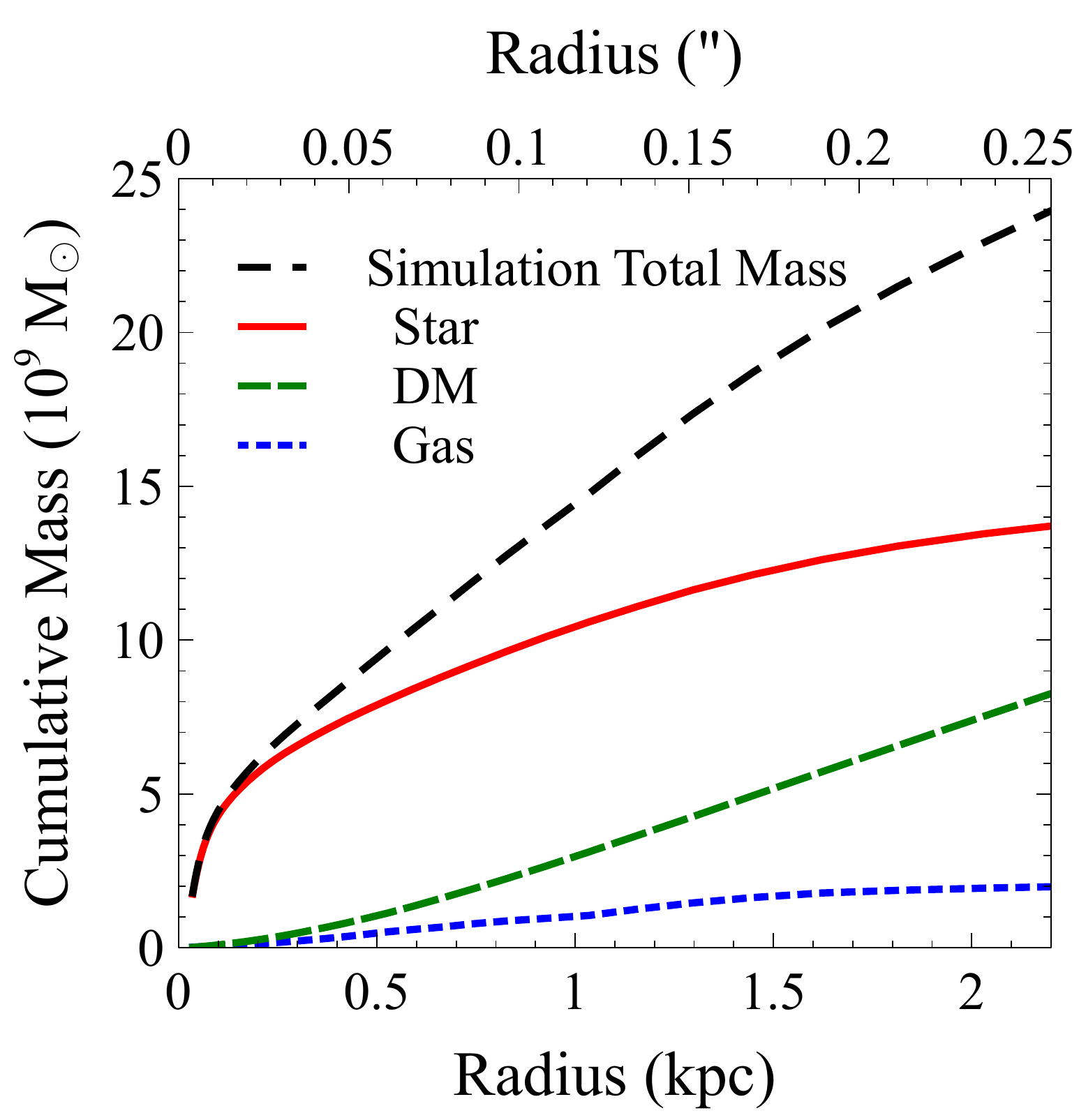}
  \includegraphics[scale=0.34]{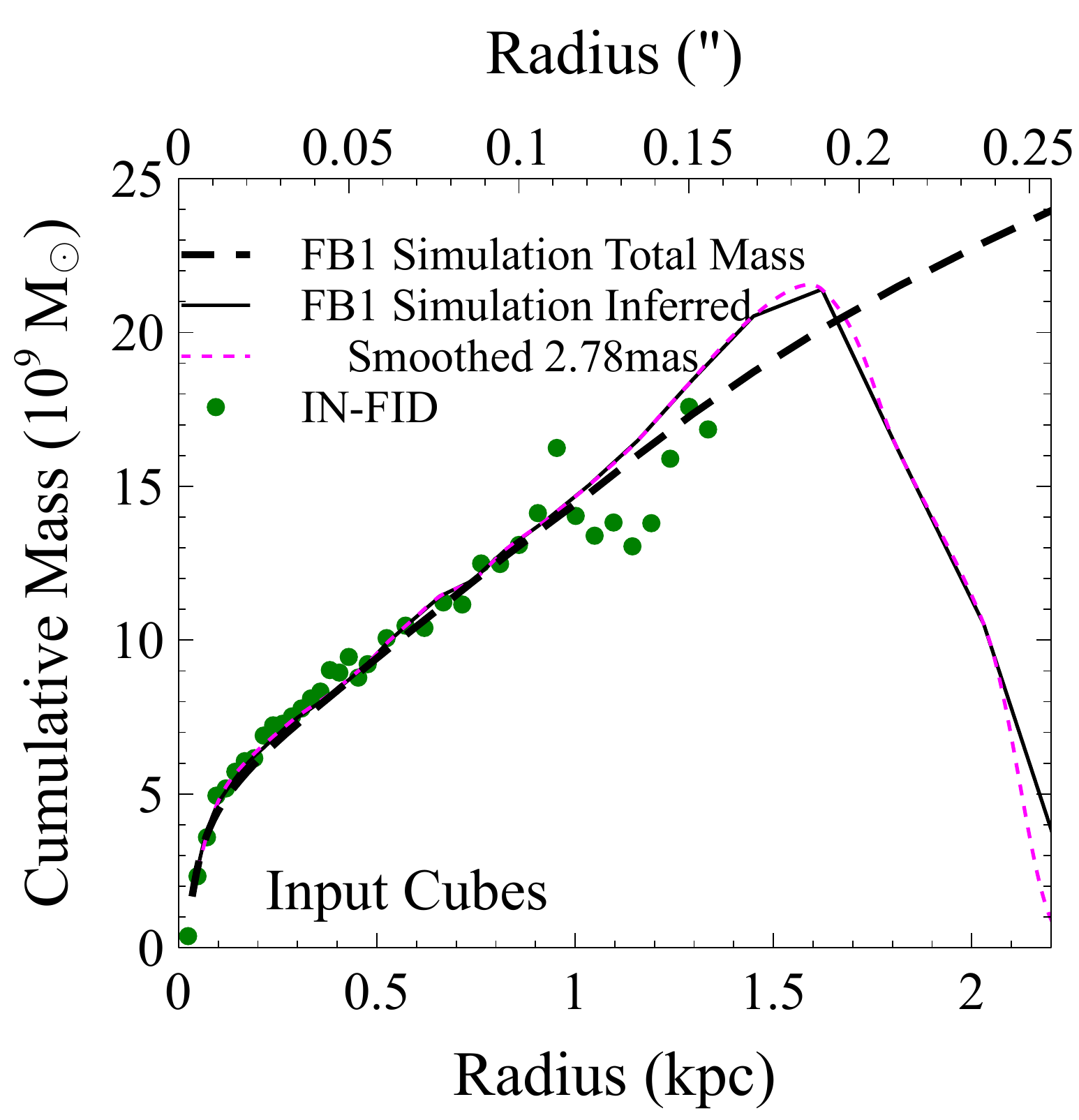}
  \includegraphics[scale=0.34]{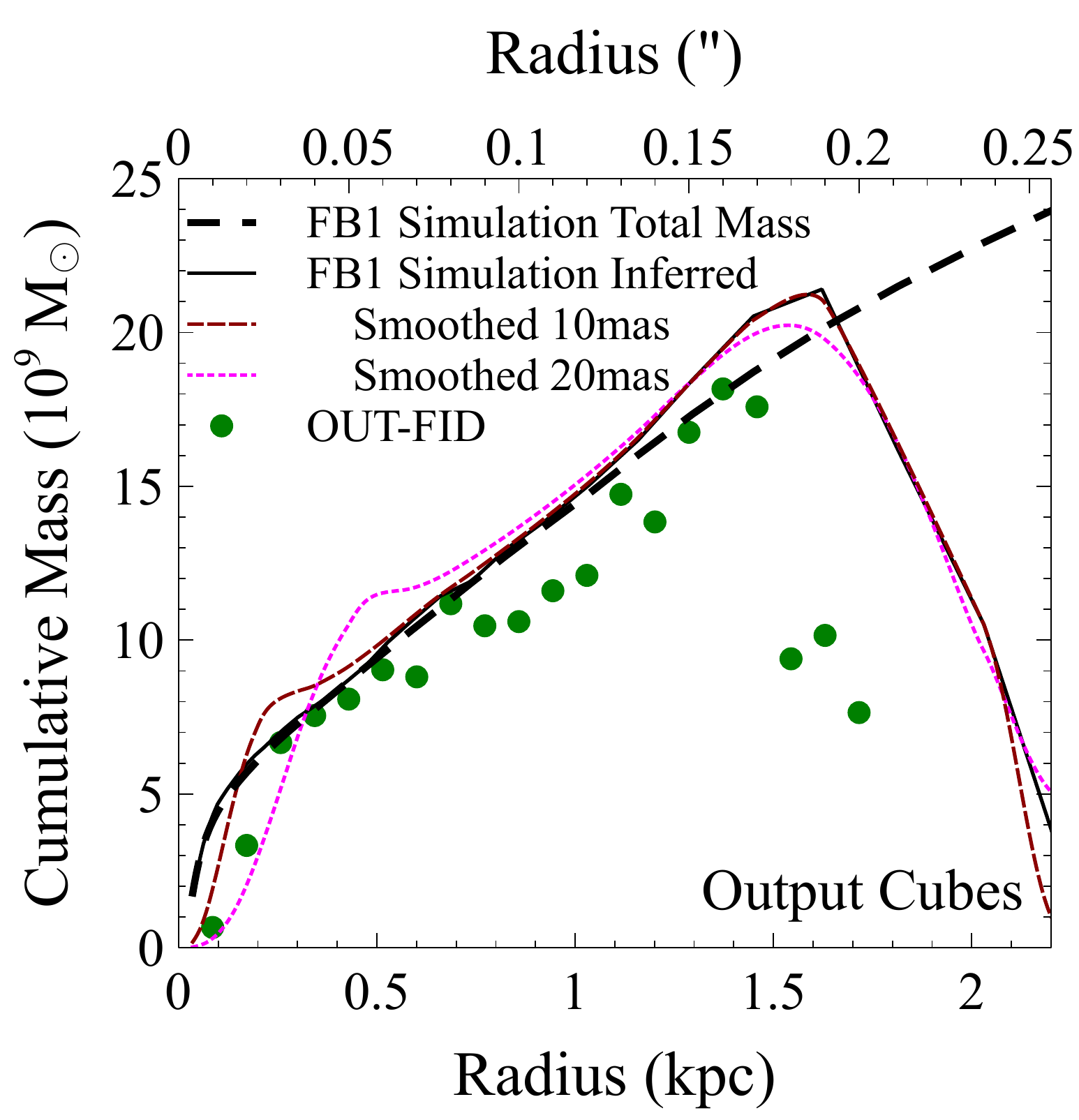}
  \caption{\small Enclosed mass vs. radius for the simulated galaxy. The mass enclosed inside a spherical volume with the specified radius is plotted for the cosmological simulation (left panel), the fiducial input cube (middle panel) and output cube (right panel). The data points show the recovered values of the dynamical mass from the mock observed rotation curves. In the middle and right panels we include the cumulative total mass from the simulation (black dashed line), as well as the inferred cumulative total mass from the simulation rotation curves that results from applying the same method as for the mock observations. The inferred rotation curves from the simulation data is smoothed to different sampling resolutions defined in the legends.}
  \label{fig:dyn_mass}
  \end{figure}

The rotation curves of galaxies can be used to determine the dynamical mass profile of the galaxies, from which we can infer the presence of dark matter \citep[e.g.,][]{Genzel17}. We use the de-projected rotational velocities of the gas, and
assuming pure rotational support of the H$\alpha$ emitting gas, we
derive the enclosed mass as a function of radius for the \SIMA\
galaxy.  The results are shown in \fig{fig:dyn_mass}. From the simulation, we compute the enclosed mass of
gas, stars and dark matter inside a spherical volume of a given
radius.  As noted in section \sect{fid}, inefficient feedback in the
\SIMA\ simulation leads to a compact bulge, with only a small
fraction (roughly 20\%) of the mass in dark matter inside the inner kpc, rising to
about a third of the total at 2\,kpc (left panel of \fig{fig:dyn_mass}).  The gas is in ordered rotation only within the inner kpc, as is evident in the middle panel, where we see small deviations from circular rotation up to truncation of the cold gas disk at 1.7\,kpc, beyond which the gas completely fails to trace the galaxy mass.

The ordered rotation of the simulated galaxy is well traced by the mock observed data,
both in the input cubes and the output cubes, up to 1\,kpc radius, as
seen in the middle and right hand panels of \fig{fig:dyn_mass}. As enclosed mass is proportional to the square of the
rotational velocity, even small deviations in the derived velocity
are amplified in the enclosed mass estimate, making the derived mass
highly sensitive to derived velocity errors. Beyond 1 kpc in the input cube, and 0.7 kpc in the output cube the velocity noise leads to a large decrement in the inferred dynamical mass. As a result, the rotation curves imply considerably less dark matter. 

We quantify the inferred dynamical masses within 0.85 kpc, half the cold gas truncation radius, inferred from the kinematics. For \OUTFID\ the inferred dynamical mass is only $9.83 \times 10^9$ $\Msun$, compared to the \INFID\ value of $12.05 \times 10^9$ $\Msun$ and real simulated value of $12.9 \times 10^9$ $\Msun$. The implication is that the fiducial observations at this radius would infer no interior dark matter. This highlights the importance of accurately recovering the rotation curves of galaxies to not underestimate the amount of dark matter, and so higher redshift observations that are not sufficiently sensitive to see the outskirts of the galaxy may suffer from these issues \citep[e.g.,][]{Price20,Tiley20}.

\subsection{Exposure time}
\begin{figure}[h]
  \centering
\includegraphics[scale=0.50]{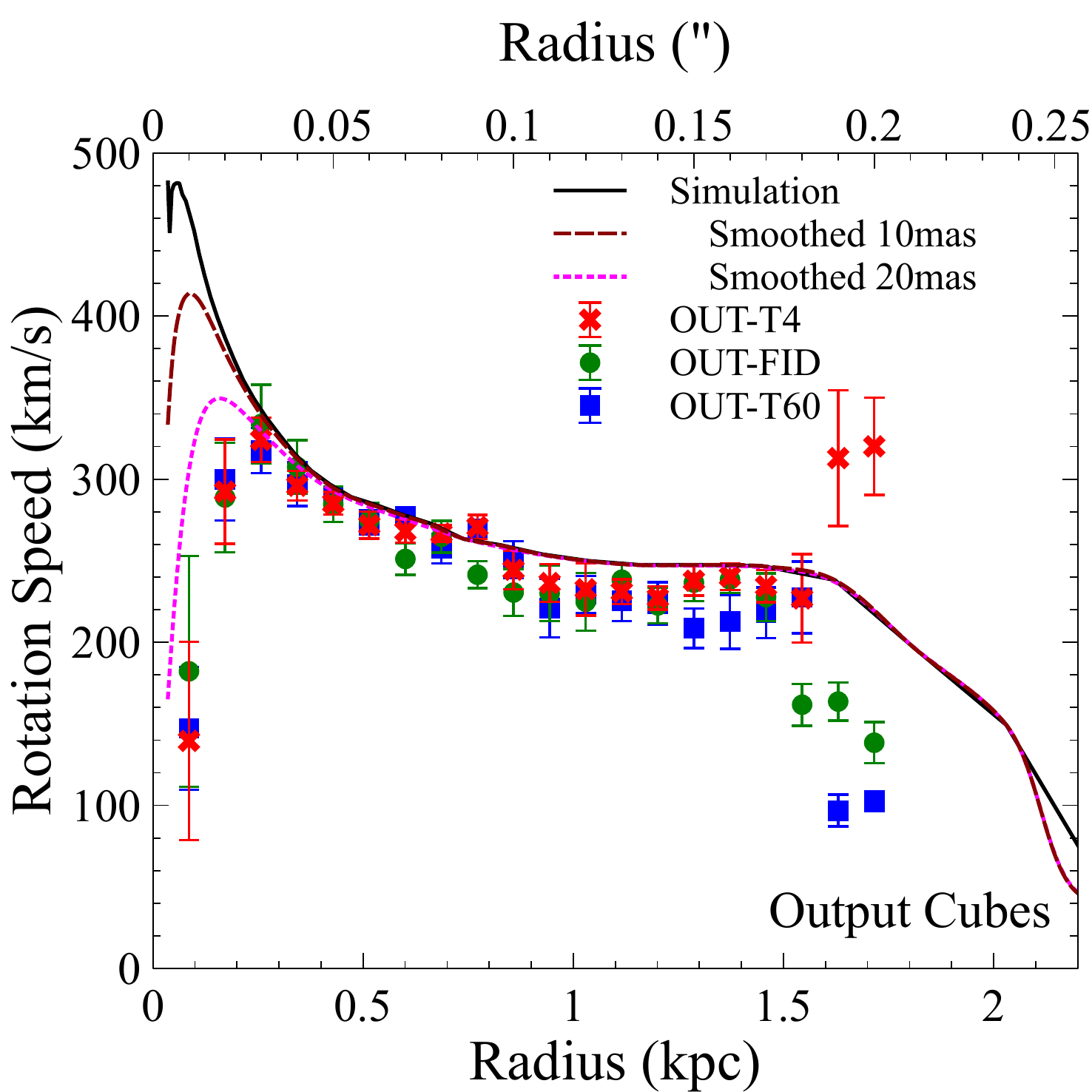}
\includegraphics[scale=0.50]{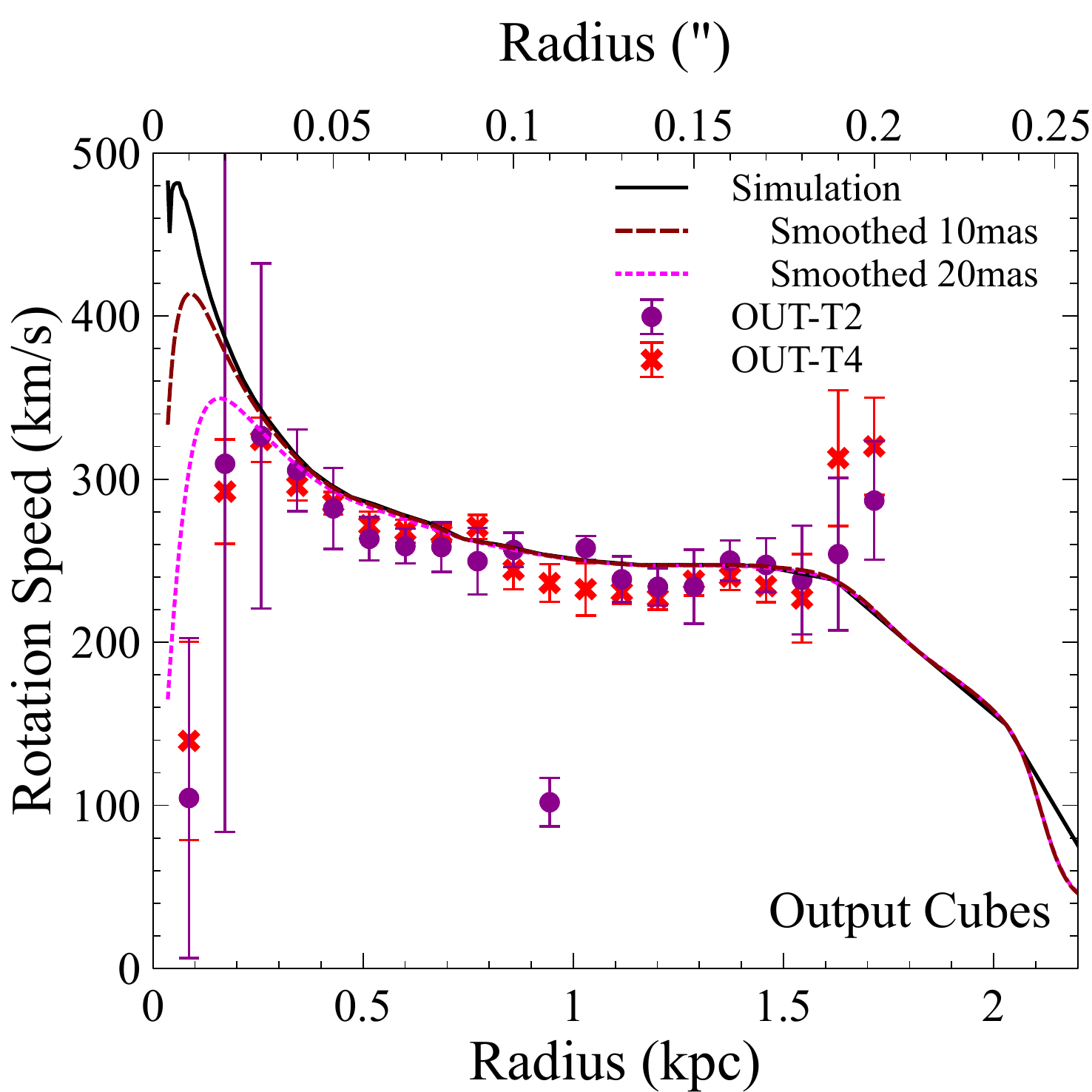}
\caption{\small Rotation curves of the galaxy comparing
  different exposure times. {\em Left}: Comparison of the fiducial (ie., 20x900s)
  with the {\tt OUT-T4} (ie, 4$\times$900s) and {\tt OUT-T60} (ie., 60$\times$900s) output cubes.  {\em Right}:
  Comparison of {\tt OUT-T2} (ie, 2$\times$900s) and {\tt OUT-T4}.}
\label{fig_time}
\end{figure}

To demonstrate the expected scaling of SNR with exposure time, as well as how the apparent rotation bias varies with SNR, we constructed three additional output cubes using the fiducial
input cube, but with 2 900s exposures ({\tt OUT-T2}), 4 900s exposures
({\tt OUT-T4}), and with 60 900s exposures ({\tt OUT-T60}). For the
reduced output cube, the median spectrally integrated SNR is 32 for {\tt OUT-T2}, 45 for {\tt OUT-T4} and 175 for {\tt OUT-T60}, all consistent
with the fiducial cube and SNR scaling as $\sqrt{t}$. We plot
the rotation curves in \fig{fig_time} for these output
cubes, where the rotation curves are sensitive to the
integrated observing time. The {\tt OUT-T2} cube has very noisy
measures of the rotation speed, as the SNR is very low. At small scales
there are so few pixels contributing to the measure of rotation that
{\tt OUT-T2} has very large uncertainties. All other output cubes are consistent with the fiducial cube, except at the largest scales where we see similar impact of more turbulent gas. Thus, as expected the exposure time is essential to accumulate an adequate SNR at small scales where there is a limited region to sample the peak rotation (c.f., \citealt{Davies11}). 

For these output cubes with varying exposure time we determine peak rotation speeds of
324 $\pm$ 13 km\,s$^{-1}$ and 318 $\pm$ 13 km\,s$^{-1}$ at 260 $\pm$ 48
pc and 254 $\pm$ 54 pc for {\tt OUT-T4} and {\tt OUT-T60},
respectively.  For the smallest exposure time investigated, {\tt OUT-T2}, we determine 327 $\pm$ 106 km\,s$^{-1}$ at 253 $\pm$ 286 pc, consistent with the peak velocity being at the centre of the galaxy.

\subsection{Dependence on Observed Spectral Resolution}
\begin{figure}[h]
\centering
\includegraphics[scale=0.50]{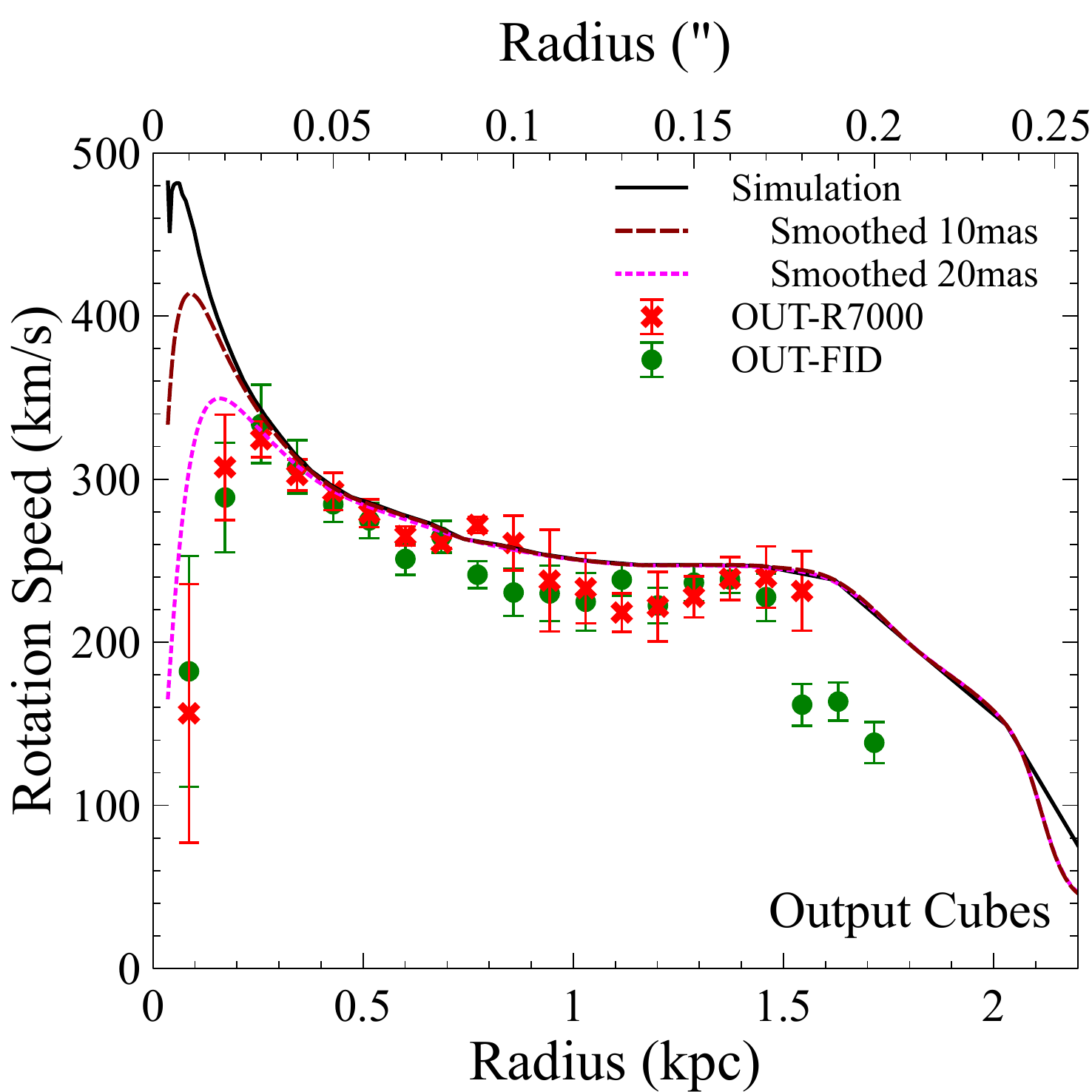}
\caption{\small Rotation curves of the galaxy taken from the 2D \diskfit\  code for fiducial spectral resolution of R$\sim$3500, and R$\sim$7000.}
\label{fig_resolution}
\end{figure}

Here we compare the impact of observing the input cube with R$\sim$3500 ({\tt OUT-FID}) and R$\sim$7000 ({\tt OUT-R7}) spectral resolving powers.
As can be seen in \fig{fig_resolution}, increasing the spectral resolution whilst holding the exposure time constant has negligible impact on the derived rotation curve, and little impact on the error bars. We recover a peak rotation speed of 325 $\pm$ 11 km\,s$^{-1}$ at $252 \pm 59$ pc, which is very consistent with the Fiducial values. This is expected, as the intensity weighted mean line-of-sight velocity observed for a spaxel is not dependent on the spectral resolving power, as long as the spectral feature is resolved, which is the case for both resolving powers chosen. In terms of signal-to-noise, doubling the resolving power reduces flux per spectral pixel to half, but also doubles the number of data points in the fit. The resulting spectrally integrated SNR is 148, $\sqrt{2}$ larger than \OUTFID, as expected. The error in the line centroid, expressed in velocity units, is only weakly dependent on the spectral resolving power. However, one difference we do see for {\tt OUT-R7} is that the higher spectral resolution leads to a more accurate recovery of the real rotation curve around 0.85\,kpc, possibly due
the wings of the spatial PSF having less impact when the spectral resolution is good. This is because the emission from other radii that gets convolved into a given spaxel causes less bias in the line of sight velocity if the velocity profile is spectrally well resolved. We discuss these sensitivity to the PSF in \sect{psf}. The dynamical mass inferred within this radius is thus the closest to the simulation value of all of our output cubes, with a value of $12.58 \times 10^9$ $\Msun$.

\subsection{Sensitivity to knowledge of the PSF}\label{psf}
The rotation curves determined from the output cubes show a small but systematic difference in their {\em flat}
part compared with the values predicted from convolving the simulation to
the appropriate spatial resolution. This could be the result of an
incorrect value for the galaxy inclination.  However, if this were the
case, the values recovered from \diskfit\ fits to the {\em input} cube
would be systematically biased, and no such bias is visible (see left
hand plot of \fig{fig_levels}). Note, too, that there is a $\sim$40\,km\,s$^{-1}$ dip in the derived rotation curve between 1.0\,kpc and 1.3\,kpc, relative to the simulation, in the input cubes. This results from the gas not being solely in ordered rotation at those radii, with a kink visible in the spiral arms (see top left corner of right plot in \fig{fig_ext}). As the effect is observed in the input cubes, it is not attributable to the instrument in any way. The galaxy exhibits a flare beyond 1.4 kpc, with a non-azimuthal velocity component over 200 km\,s$^{-1}$. When correcting for inclination and convolving with the local rotation speed, the inferred rotation is roughly 265 km\,s$^{-1}$, consistent with the higher dispersion, fast rotation seen at these large scales.
We note that the simulated data allow us to have much more accurate knowledge of
the inclination to the line-of-sight, in contrast to a real
observation, where the inclination is inferred by fitting an ellipse
to the galaxy image, either broadband (continuum) or narrow-band
(emission line), and assuming an intrinsically round geometry.  A systematic error
in the inclination 
$i=51^\circ$ 
leads to a
bias in the inferred rotation
velocity and the dynamical mass estimates. 

\begin{figure}[h]
\centering
\includegraphics[width=0.35\textwidth, trim=25 0 50 0,clip]{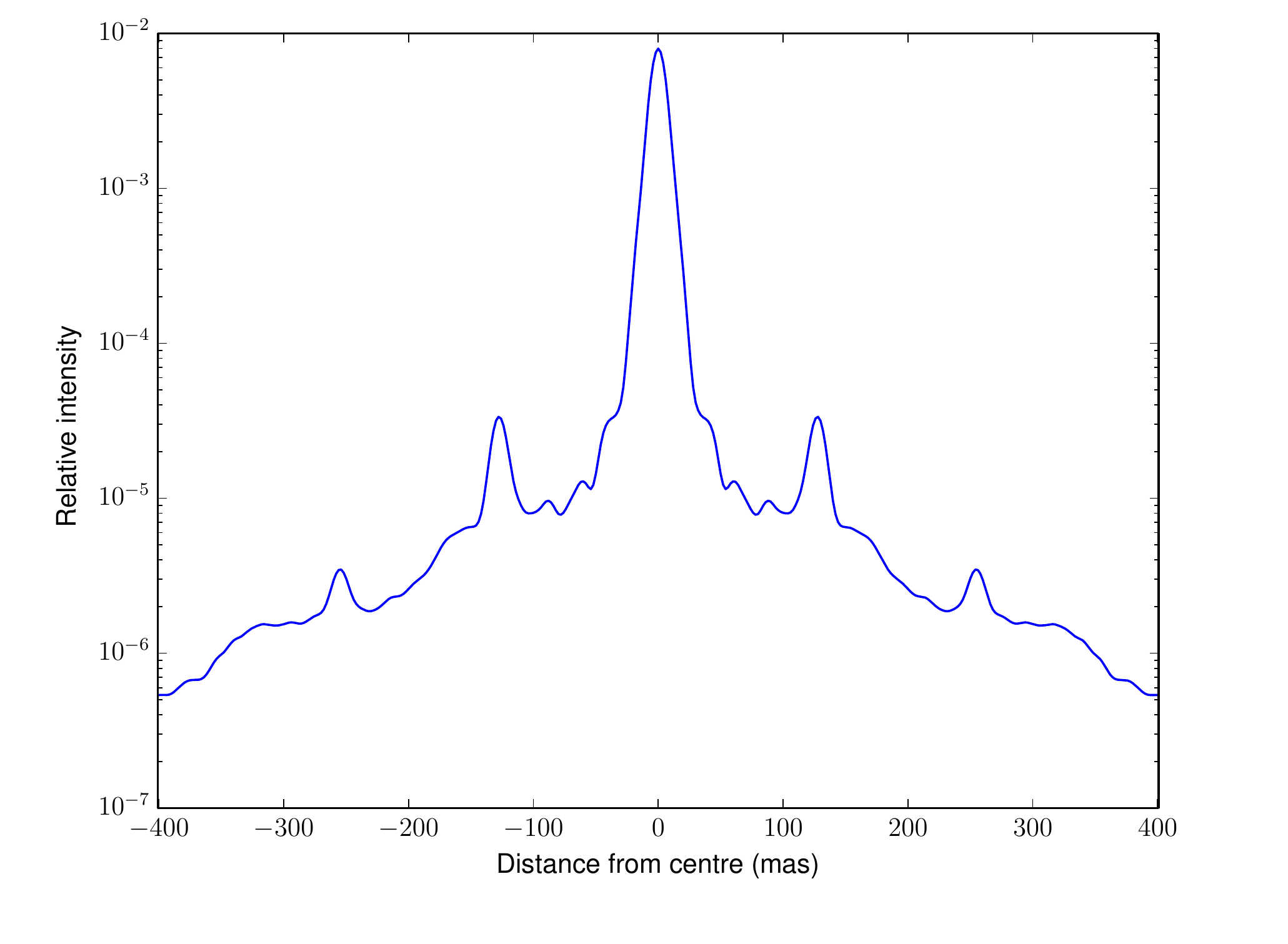}
\includegraphics[width=0.296\textwidth, trim=34 0 108 0,clip]{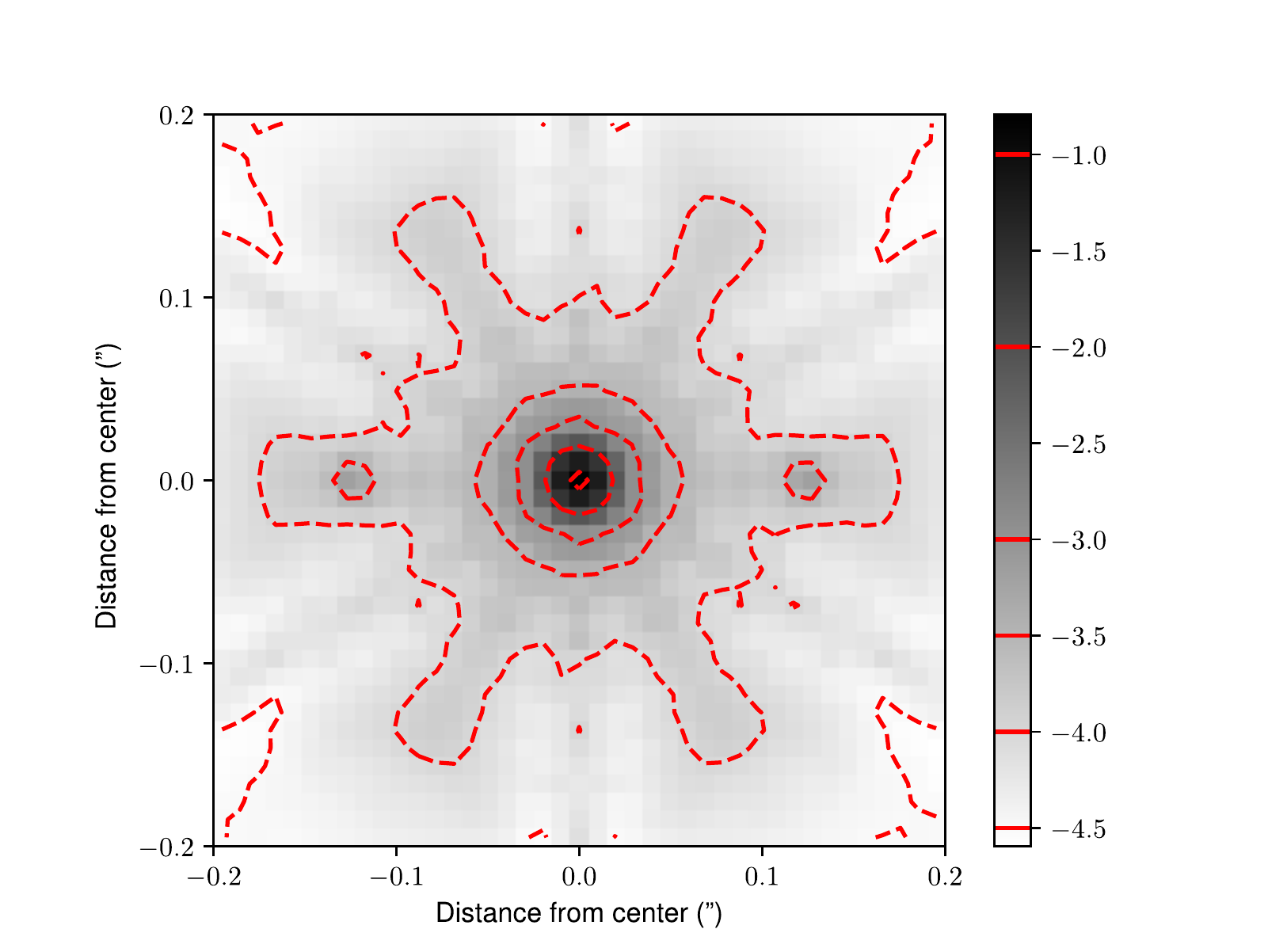}
\includegraphics[width=0.33\textwidth, trim=50 0 55 0,clip]{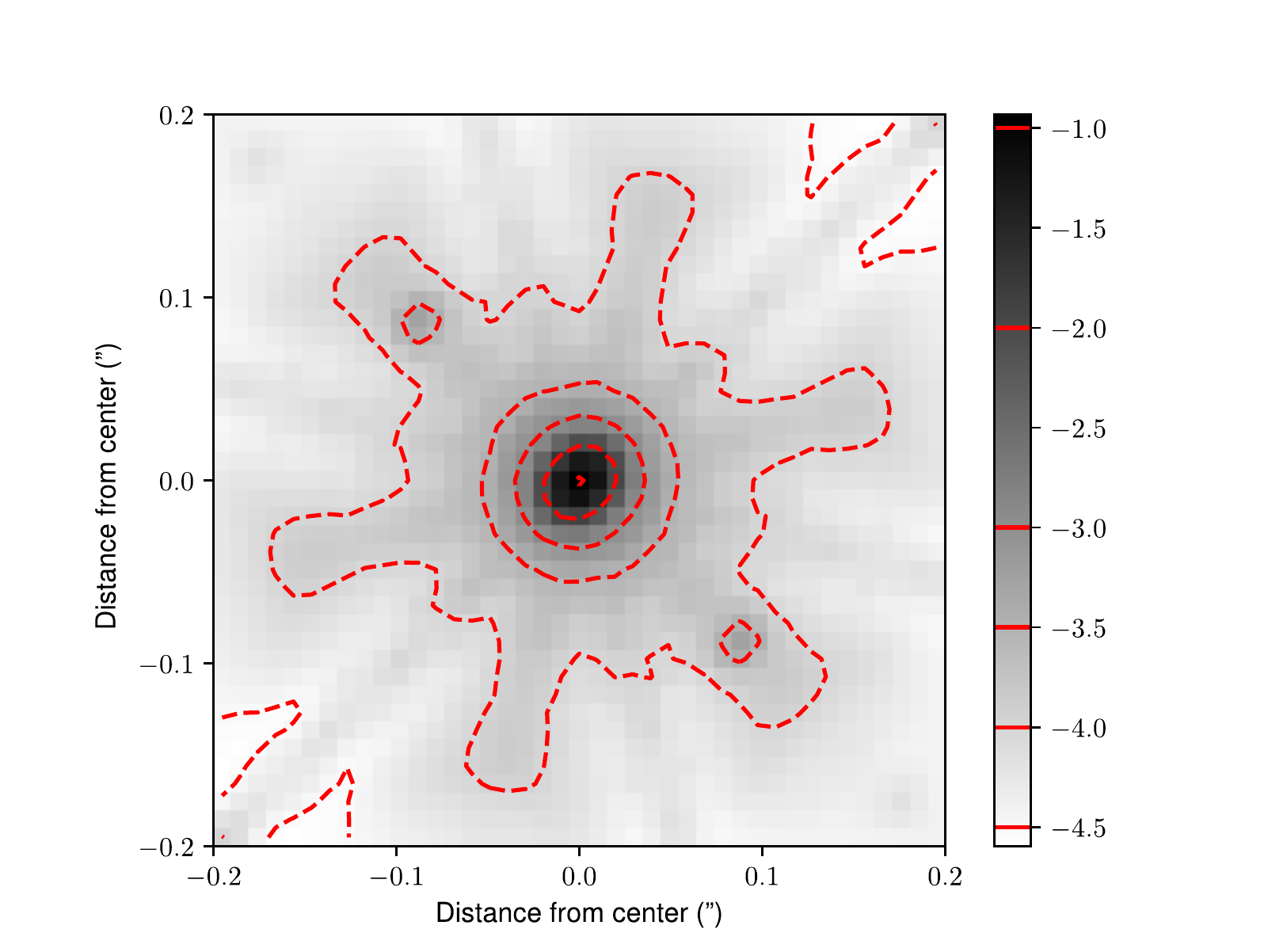}
\caption{\small 2D and 1D plots of typical HARMONI LTAO PSFs. [Left] 1D plot of azimuthally averaged intensity vs. radius for the PSF of the middle panel, showing the sharp core and extended wings. [Middle] 2D log plot of the central 0{\mbox{\ensuremath{.\!\!^{\prime\prime}}}}4$\times$0{\mbox{\ensuremath{.\!\!^{\prime\prime}}}}4 of the LTAO PSF at 1.6\,${\rm \mu}$m with 2\,mas $\sigma$ residual jitter along both axes. The contours levels are indicated on the log intensity scale bar. Note the 6-pronged low level contours arising from the 6 LGS.  [Right] Same as the middle panel but with 2\,mas and 6\,mas $\sigma$ residual jitter, with the larger jitter axis at 45$^\circ$.}
\label{fig_psf}
\end{figure}

The systematic
lowering of observed rotation values in the output cube could be due
to the wings of the AO PSF, as the AO only achieves partial correction
of the atmospheric turbulence (see \fig{fig_psf}).  As these wings extend out to large
radii ($\sim$ 100--500 mas FWHM for median seeing at H band), they could
have a substantial effect on the observed kinematics.

\begin{figure}[p]
\centering
\includegraphics[width=0.48\textwidth]{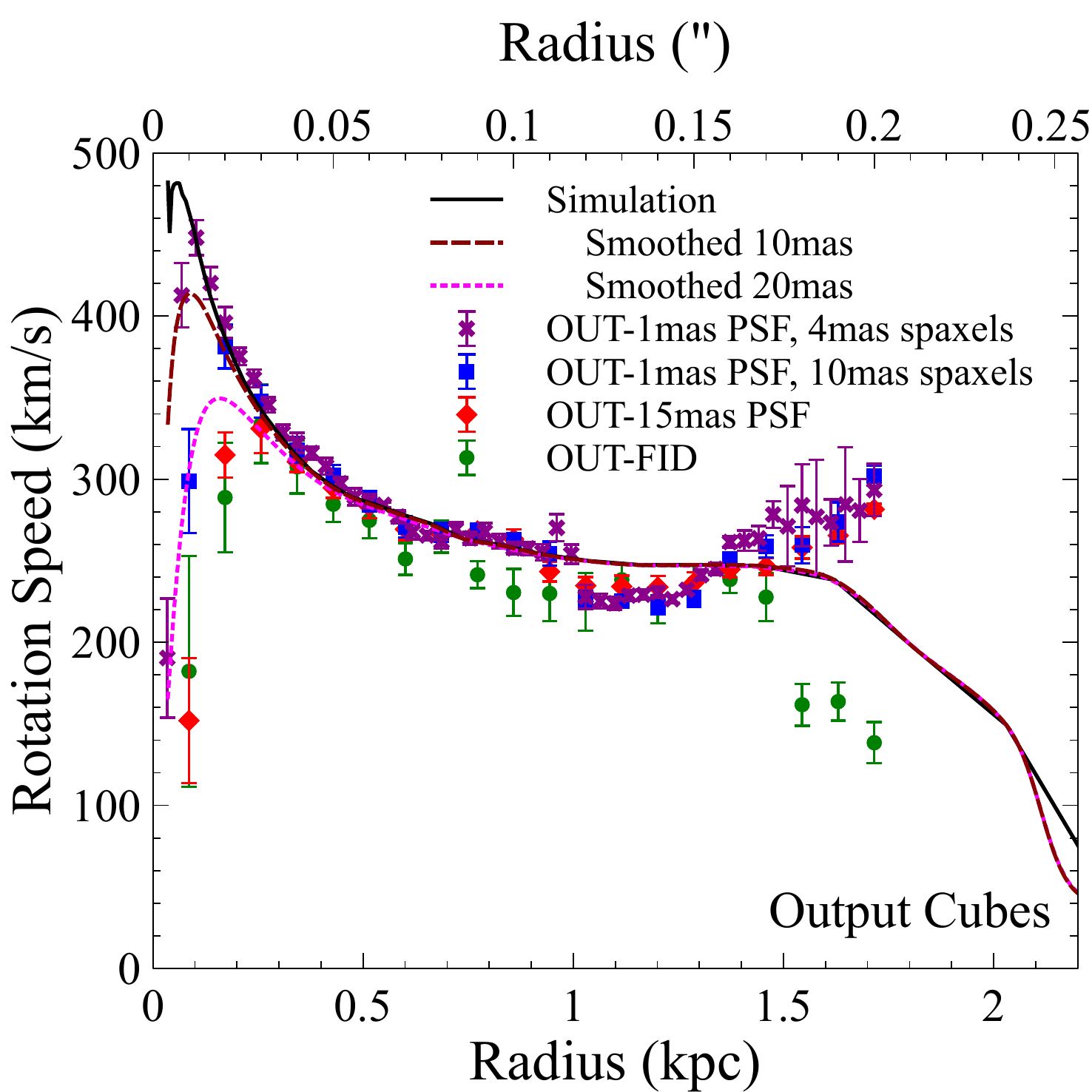}
\includegraphics[width=0.48\textwidth]{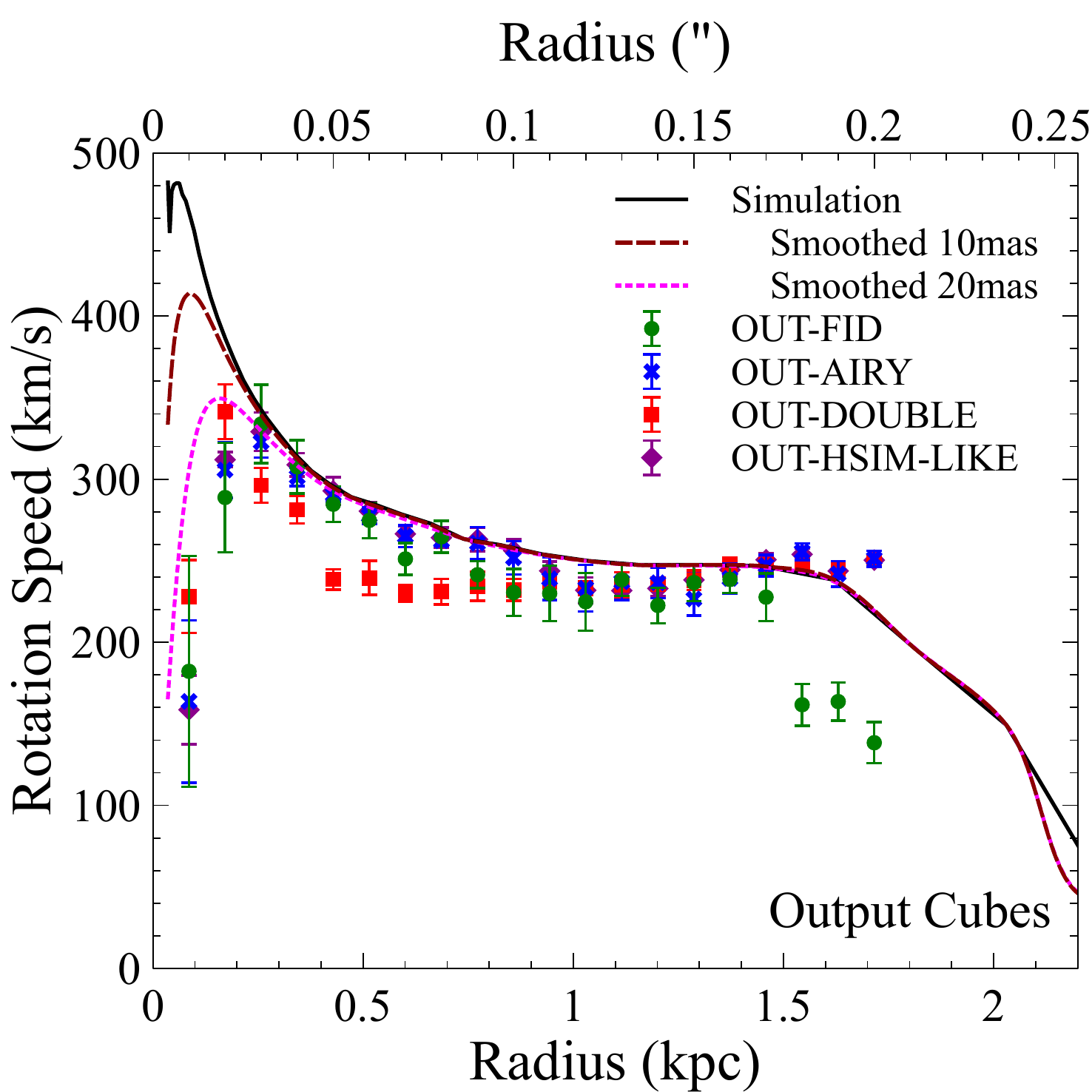}
\includegraphics[width=0.48\textwidth]{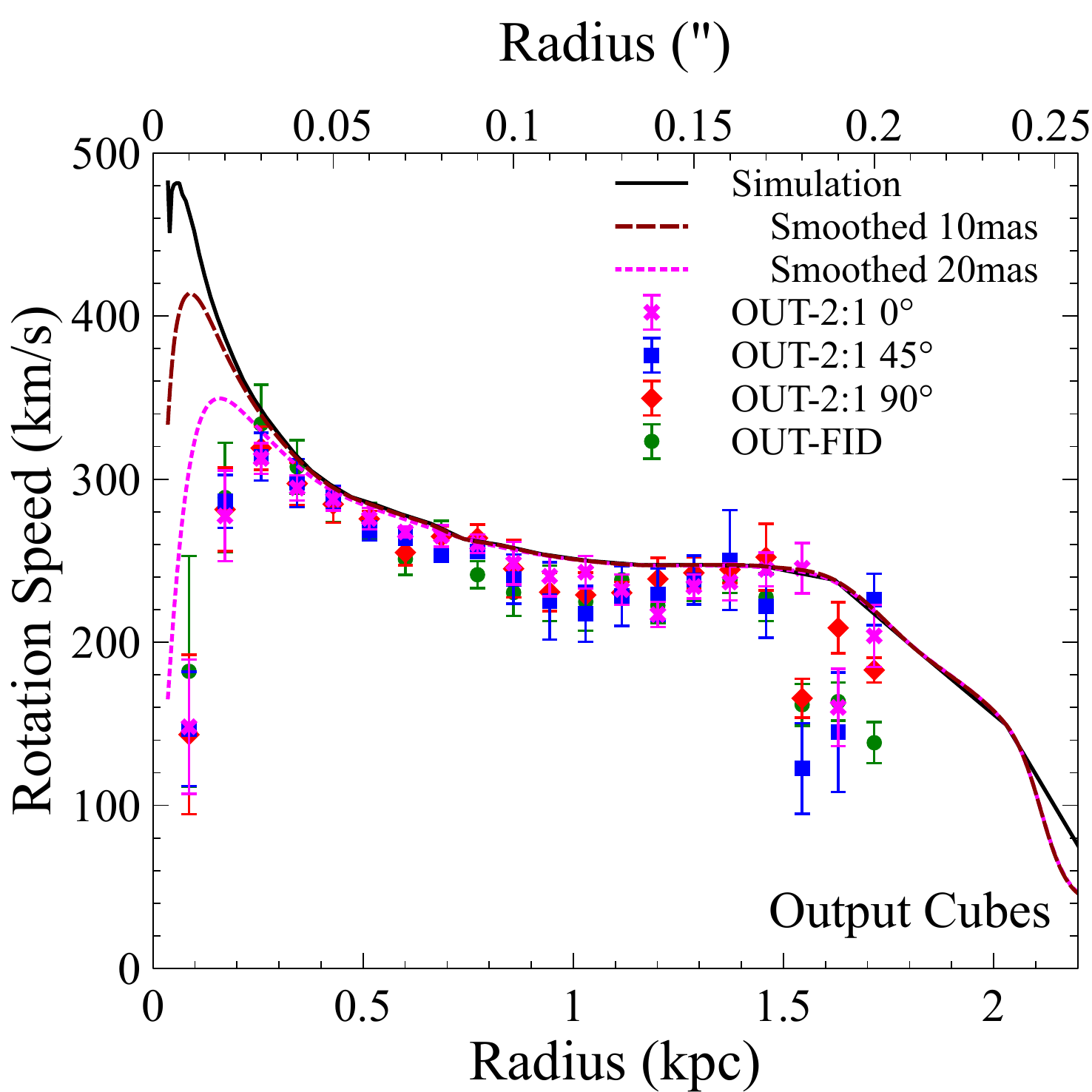}
\includegraphics[width=0.48\textwidth]{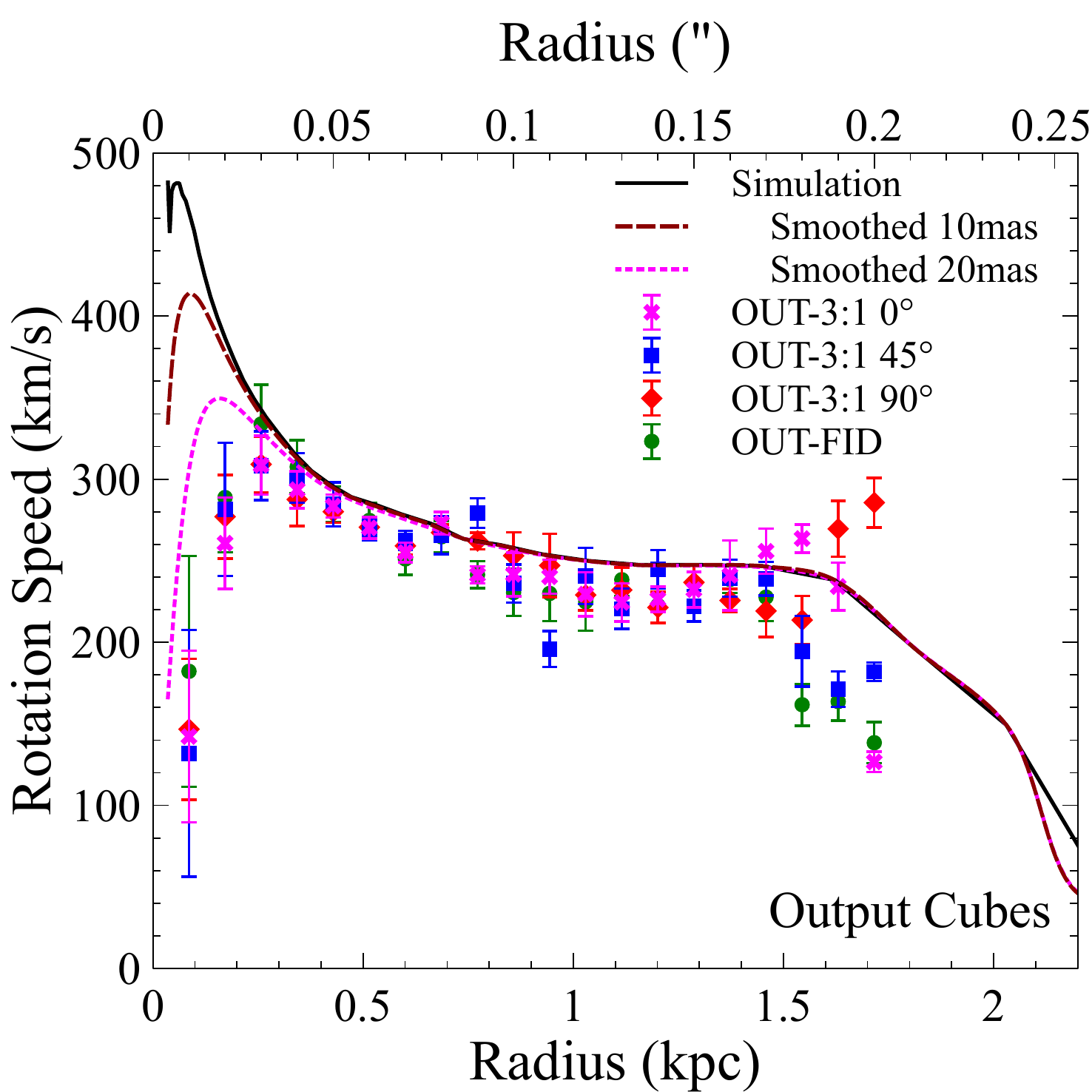}
\caption{\small Rotation curves of the galaxy with instrumental PSFs that are
 [Top-left] pure Gaussians with
  1\,mas and 15\,mas FWHM, the former with fine and coarse spaxels.
[Top-right] Comparison with a pure Airy pattern (with central
obscuration), and a two Gaussian component PSF with 15 mas
and 100 mas FWHM, with the same flux ratio as the LTAO PSF.  [Bottom-Left] Elongated LTAO PSFs with 4.7$\times$9.4 mas
(FWHM) residual tip-tilt jitter at three different angles and [Bottom-right] Elongated LTAO PSF with 4.7$\times$14 mas FWHM tip-tilt jitter at three different angles. In each case, the derived
rotation curve is compared with \OUTFID.  Long exposure times were used to discount SNR effects.}
\label{fig_airy}
\end{figure}

To test whether the wings of the PSF were responsible for reducing the recovered rotation speed,
we re-ran the \hsim\ simulation with several different artificial
PSFs. Using a 1\,mas Gaussian, with 4\,mas spaxels exactly reproduces
the input cube, showing that there is no systematic impact from using
\hsim.  A 1\,mas PSF with 10\,mas spaxels shows negligible difference in the flat part of the rotation curve (between 0.4 and 1\,kpc), demonstrating that sub-Nyquist sampling is not an issue (top left panel of \fig{fig_airy}). A single 15\,mas FWHM Gaussian and a pure Airy PSF with 10\,mas spaxels (describing the telescope as a circular aperture with a
circular central obscuration) both mimic the 1\,mas PSF, so the 15\,mas PSF width does not cause the systematic lowering either.  Note that the coarse sampling (10\,mas spaxels) and the 15\,mas width both substantially impact the steepness of the rotation curve's rise in the innermost part, as would be expected due to the blurring caused by limited spatial resolution.

We also explored two-component PSFs (top right panel of
\fig{fig_airy}). The results for 1\,mas--100\,mas and a
15\,mas--88\,mas double Gaussian PSFs, both with flux ratios for the two components similar
to the LTAO PSF, show that the extended component, which folds in
kinematic information from large radii, is mostly responsible for the
systematic lowering of the rotation curve in its flat part. Together
with the effect of the sampling apparent in the innermost parts of the rotation curve, we can fully account for the observed
differences. 

Another typical AO PSF artifact is an elongation of the PSF along the
direction towards the AO tip-tilt natural star (the star is up to
60$^{\prime\prime}$ off-axis for HARMONI LTAO). This is the result of a
de-correlation of atmospheric tip-tilt with off-axis field angle.  To
mimic the effects of an elongated PSF, we re-ran \HSIM\ with three
different residual tip-tilt jitter values -- no jitter, Gaussian
residual jitter with $\sigma$ of 2 mas along $x$ and 4 mas along
$y$-axis of the image, and Gaussian residual jitter with $\sigma$ of 2
mas along $x$ and 6 mas along $y$ (right panel of \fig{fig_psf}). The no jitter situation is
un-physical, but provides a reference point for comparison,
particularly with the pure Airy PSF. The two elongated PSFs with 2:1 and 3:1 aspect ratios have been simulated at three different position angles of 0$^\circ$, 45$^\circ$ and 90$^\circ$.

In comparison to the \OUTFID\ simulation, the elongated PSFs result in systematically lower values of rotation for the innermost regions, where the rotation curve is steeply rising. This is expected as the elongation results in values at different radii getting mixed together by the PSF convolution. In the flat part of the rotation curve, both elongated PSFs at 45$^\circ$ angle show lower rotation values than those at 0$^\circ$ and 90$^\circ$. The discrepancy is most noticeable around 1\,kpc radius. Otherwise, there is no discernible difference between the elongated PSFs and the round one.

\section{Summary and Conclusions}\label{conclusions}
\subsection{Summary} To aid in predicting the
detailed morphology, kinematics and dynamics of high-redshift galaxies
as observed in gas emission lines, we have incorporated a method of computing the H$\alpha$ emission from every cell in the
{\tt RAMSES} AMR cosmological simulation of a galaxy at $z_{\rm f}=3$. The method uses the instantaneous star formation rate in the cell, together with the
Kennicutt-Schmidt relation (K94) to predict the recombination line flux from
each cell.  The flux computation is combined with
known kinematics and estimated turbulence of each gas cell,
together with estimates of the extinction along a chosen line-of-sight,
to yield a {\em data cube} of emission line intensity as a function of
two spatial coordinates (the galaxy as viewed in the plane of the sky)
and one wavelength (or equivalently, line-of-sight velocity)
coordinate. This pipeline, \rtoh, is made available to the community.

We then simulate observations of this {\em mock} galaxy, artificially
redshifted to $z=1.44$, made with the HARMONI integral field
spectrograph on the ELT in the H+K band (at 1.6\,$\mu {\rm m}$).  The
simulated observations account for a variety of atmospheric and
instrumental effects, including sky emission and absorption, shot
noise from thermal background and sky lines, detector dark and read
noise, and the adaptive optics and instrumental PSFs.  The {\em observed}
data cube is then analysed to recover the kinematic and dynamical
information about the galaxy in a spatially resolved manner (i.e. the galaxy's rotation curve), and this is compared with the {\em
  original} kinematics derived from the simulation input, thus highlighting any biases arising from the observation. We vary the sampling, exposure time, and spectral resolution of the observation to study their impact on recovering the galaxy characteristics.

\subsection{Conclusions}
There is no evidence of bias in the rotation curve derived from the
input cubes, prior to observation with HARMONI, when compared with the
rotation velocities of the gas deduced from the simulation data. Thus, we can be
confident that we have a correct method of translating the
cosmological simulations into emission line kinematics. However, the rotation curve derived from the input cube deviates from the dynamical value at radii where a substantial fraction of the line emitting gas is not in ordered rotation, as seen in \fig{fig_levels} between 1 and 1.3\,kpc radius.


A full kinematic fit with \diskfit\ using the two dimensional line-of-sight velocity
information from integral field spectroscopy recovers a rotation curve closely
aligned with the native simulation values, when the latter is convolved to the spatial
resolution of the mock observations.

Adequate signal to noise ($\sim$7 per spaxel) integrated across the
spectral line is required for a robust extraction of the galaxy's
rotation curve. We find that when the SNR drops below this value (60
min integrated exposure time), we
increase the uncertainty in the ability to trace the rotation curve at all
radii (as shown for the {\tt 2$\times$900s} exposure).
This increased uncertainty for measured rotation in both the inner and outer regions is expected; at small radii
there are few data points to average, so there is strong dependence on
the SNR.  At large radii ($\geq$1.5 kpc), the galaxy displays
strong departures from ordered disk rotation, in particular there
appears to be a warp, and tidally stripped gas from previously
accreting substructure with markedly different kinematics.  At these
radii, emission is also weaker, leading to poor SNR and large
uncertainties in V$_{\rm rot}$.


The simulations show that HARMONI is able to observe the rotation
curve of a 3\,$\Msun$ ${\rm yr^{-1}}$ star forming galaxy (1.25\,L$_*$ at
$z = 1.44$) in 60 minutes of total exposure time.
We are able to mostly recover the
enclosed mass profile via dynamical mass measurements, though the
uncertainties in the measured quantities need to be propagated correctly to the derived parameters, given the strong
dependence of derived mass on observed line-of-sight velocity.

The resulting rotation curves 
show that the H$\alpha$ is a
good tracer of the galaxy kinematics, as borne out by the excellent
match with the (smoothed) simulation input.  As discussed in detail in other works, we see strong beam-smearing effects which smooth the rotation curves in all cases, making
them appear flatter, and suppressing the central peak \citep[e.g.,][]{Davies11,Genzel14,Burkert16,Tiley19}. The latter is
expected from the reduced spatial resolution of the observed
cube, but the recovered rotation curve is even flatter than the input
simulation smoothed to 20\,mas resolution, particularly in the region
where the rotation curve turns over from rising to flat. 

A key result is that the instrument and AO PSF has a strong impact on the observed rotation
curve.  We find that the two-component nature of the LTAO PSF
(diffraction-limited core and extended wings of
$approx$ 100--500\,mas) causes a systematic
lowering of the measured values of the rotation curve in its {\em flat} part, also impacting the derived dynamical mass. This lowering of the rotation curve is not as significant when using a higher spectral resolution. Coarse (sub-Nyquist) sampling of the output has a negligible impact in the {\em flat} part, but does heavily influence the fast rise in the inner part of the rotation curve, smoothing the steep rise. The effective spatial
resolution of the observation appears to be limited to being twice the
spaxel scale, as might be expected from the Nyquist sampling theorem,
and similar to previous results \cite[e.g.,][]{Evans15, Puech16, Puech18}. 
The recovered velocity profile is sensitive to asymmetries in the PSF
shape. Strong asymmetries in the residual tip-tilt jitter (3:1
aspect ratio), particularly when the PSF elongation is not aligned with the axes of the integral field unit, make it difficult to extract a robust rotation curve.
A significant
conclusion of this work is that good instrument and AO PSF knowledge is
required to correctly estimate the bias introduced in measurement of
galaxy physical parameters.

Finally, the spatial resolution of the input simulation has little
impact on the rotation curve recovered from the mock
observations -- the only discernible effect is the slight {\em smoothing} of
the kinematics in the input cube.  Even the coarsest cell size used in our simulations is
still only half the size of the HARMONI spaxels, so we can usefully
limit the spatial resolution of the simulations to $\sim$5\,mas cell
sizes without any impact. This result will be useful for future
simulation runs.

\section{Acknowledgements}

We would like to thank members of the HARMONI science team for useful
discussions about the simulations and the content of the paper.  We
are particularly indebted to Simon Zieleniewski, author of the
original \HSIM\ code, on which this research is based.  
This work used the DiRAC Complexity system, operated by the University of Leicester IT Services, which forms part of the STFC DiRAC HPC Facility (www.dirac.ac.uk). This equipment is funded by BIS National E-Infrastructure capital grant ST/K000373/1 and  STFC DiRAC Operations grant ST/K0003259/1. DiRAC is part of the National E-Infrastructure.
MLAR would like to thank Adrianne Slyz and Julien Devriendt for useful conversations and mentorship that contributed to this work. The research of MLAR was partially
supported by Adrian Beecroft. MPS acknowledges support from the Comunidad de Madrid through Atracci\'on de Talento Investigador Grant 2018-T1/TIC-11035. We
also thank the referee for very useful and insightful comments that helped make this manuscript more focused.  Research
for this paper was supported by a grant from the Science and
Technology Facilities Council (part of UKRI) to the University of
Oxford as part of the ELT Programme.  The authors
acknowledge support from grants UKRI grants ST/N002717/1, ST/M007650/1
and ST/S001409/1.

\section*{Data Availability}
The \rtoh\ pipeline described in this manuscript is made available without any guarantee of usability. The software can be found here: \href{https://github.com/mlarichardson/ramses2hsim}{https://github.com/mlarichardson/ramses2hsim}.
\bibliography{references.bib}

\end{document}